\documentclass[epj]{svjour}
\usepackage{graphicx}
\usepackage{amsmath}

\journalname{Eur. Phys. J. A}

\usepackage{epstopdf}
\def\be{\begin{equation}}
\def\ee{\end{equation}}
\def\bea{\begin{eqnarray}}
\def\eea{\end{eqnarray}}
\def\ba{\begin{array}}
\def\ea{\end{array}}
\def\bc{\begin{center}}
\def\ec{\end{center}}
\newcommand{\bfq}{{\bf q}_{\perp}}

\newcommand{\bfp}{{\bf p}_{\perp}}

\usepackage{color}

 \newcommand{\ud}{\text{d}}
\newcommand{\mb}[1]{\boldsymbol{#1}}

\begin{document}

\title{Gravitational form factors and angular momentum densities in light-front quark-diquark model}
\author{Narinder Kumar$^1$\and Chandan Mondal$^2$ \and Neetika Sharma$^{3,4}$}

\institute{Department of Physics, Indian Institute of Technology Kanpur, Kanpur-208016, India.\and Institute of Modern Physics, Chinese Academy of Sciences, Lanzhou-730000, China.\and Department of Physical Sciences,
I K Gujral Punjab Technical University, Jalandhar 144603, Punjab, India.\and
Department of Physics, Panjab University, Chandigarh 160014, India.}


\date{Received: date / Revised version: date}

\abstract{We investigate the gravitational form factors (GFFs) and the longitudinal momentum densities ($p^+$  densities) for proton in a light-front quark-diquark model. The light-front wave functions are constructed from the soft-wall AdS/QCD prediction. The contributions from both the scalar and the axial vector diquarks are considered here. The results are compared with the consequences of a parametrization of nucleon generalized parton distributions (GPDs) in the light of recent MRST measurements of parton distribution functions (PDFs) and a soft-wall AdS/QCD model. The spatial distribution of angular momentum for up and down quarks inside the nucleon has been presented. At the density level, we illustrate different definitions of angular momentum explicitly for an up and down quark in the light-front quark-diquark model inspired by AdS/QCD. 
} 
\authorrunning{N. Kumar \and C. Mondal\and N. Sharma}
\titlerunning{GFFs and angular momentum densities in light-front quark-diquark model}
\maketitle
\vskip0.2in
\noindent
\section{Introduction}
Nucleon tomography has become an important tool in the modern study of the nucleon structure \cite{Diehl2003,Goeke2001}. One can characterize the distribution of quarks in high energy nucleon not only by the momentum fraction $x$  but also by the transverse position $\vec{b}_\perp$ and transverse momentum $\vec{k}_\perp$. This important information is encoded in the generalized parton distributions (GPDs) and transverse momentum distributions (TMDs). We can access the GPDs in hard exclusive processes like deep virtual compton scattering (DVCS) or deep virtual meson production (DVMP) and TMDs in the semi-inclusive processes like semi-inclusive deep inelastic scattering (SIDIS). Worldwide, several experiments such as, the H1 collaboration \cite{Adloffetal.2000a,adloff}, ZEUS collaboration \cite{Breitwegetal.1999,Chekanov2003} and fixed target experiments at HERMES \cite{Airapetian2001} have finished taking data on DVCS. Experiments are also being done at JLAB, Hall A and B \cite{Stepanyan2001} and COMPASS at CERN \cite{dHoseN.2004} to access GPDs and new data is expected from the upcoming experimental facilities at JLAB12 and electron ion collider (EIC) \cite{AccardiA.2016}.

In literature, several models \cite{Pasquini2005b,Pasquini2006,Dahiya2008a,Frederico2009,Mukherjee2013,Maji:2015vsa} and parametrizations \cite{Goldstein2011a,Goldstein2015,neetika2016} are present for GPDs. The Fourier transform of GPDs with skewness equal to zero gives the impact parameter dependent parton distribution function (ipdpdf) providing the information of the partons of a given longitudinal momentum distributed in transverse position space. The $x$ moments of GPDs give the form factors accessible in exclusive processes whereas in the forward limit they reduce to parton distributions, accessible in inclusive processes. Electromagnetic form factors (EFFs) \cite{Sachs1962} describe the spatial distributions of electric charge and magnetization densities inside the nucleon and thus are intimately related to its internal structure; these form factors are among the most basic observables of the nucleon. The Fourier transform of EFFs gives the charge and magnetization distributions of nucleon respectively.
One can obtain the Dirac, $F_1(Q^2)$ and Pauli, $F_2(Q^2)$ form factors from the first moment of spin non-flip $H(x,Q^2)$ and spin flip $E(x,Q^2)$ GPDs \cite{Radyushkin1997}.

Gravity plays a major role at two extreme but completely different scales i.e., Planck and cosmic. At the subatomic levels, gravity has little effect due to its weak coupling. In classical mechanics, mass distribution and moment of inertia are important concepts but barely discussed for quantum systems like nucleon. The second moment of charge distribution gives the mass distribution for nucleon. However, it is very interesting that the second Mellin moments GPDs give the gravitational form factors (GFFs) without actual gravitational scattering. In hadron physics, matrix elements of the energy-momentum tensor $(T^{\mu \nu})$ relate to the GFFs $A(Q^2)$ and $B(Q^2)$ \cite{Brodsky2001}. GFFs $A(Q^2)$ and $B(Q^2)$ can be obtained from the helicity non-flip and helicity flip matrix elements of the $T^{++}$ component. The helicity non-flip GFF $A(Q^2)$ provide us the momentum fraction carried by the each constituent of a hadron whereas $ B(Q^2)$ gives the value of the gravitomagnetic moment at $Q^2=0$ which is in agreement with equivalence principle and energy-momentum conservation \cite{Kobzarev1962,Kobsarev1970}. The GFFs $A(Q^2)$ and $B(Q^2)$ are also related to the second moment of spin non-flip $H(x,Q^2)$ and spin flip $E(x,Q^2)$ GPDs. 
It is important to study the GFFs as they provide information on the nucleon spin, according to Ji's sum rule: $2 \langle J_q \rangle = A_q (0) + B_q (0)$ \cite{Ji1997a,Ji1998}. GFFs have been studied in different models, e.g., light-front QED \cite{kumar2014a,Brodsky2001}, Anti-de Sitter/Quantum Chromodynamics (AdS/QCD) models \cite{Brodsky2008,Abidin2009,Abidin2008a,Abidin2008b,Mondal2016a}, light-front quark-diquark model\cite{Chakrabarti2015}, phenomenological parametrization \cite{Selyugin2009} etc.. It is now well established that a correspondence exists between the transition amplitudes describing the interaction between string modes in AdS space and matrix elements of the $T^{\mu \nu}$ tensor of the fundamental hadronic constituents in QCD \cite{Brodsky2008}. Gravitational form factor $A(Q^2)$ for nucleon in both hard-wall and soft-wall AdS/QCD model have been calculated in \cite{Abidin2009}. In Ref. \cite{Abidin2008a}, pion and axial-vector mesons in hard-wall AdS/QCD model have been studied, whereas GFFs in holographic model of QCD for vector mesons have been studied in \cite{Abidin2008b}.
Structure of transverse polarization of nucleon from energy-momentum tensor is also explained by Ji \textit{et. al} \cite{Ji2012}, where they have considered the partonic contributions from the leading, sub-leading and next to sub-leading parts in the light-cone coordinates. 
GFFs  $A(Q^2)$, $B(Q^2)$ and $\bar{C}(Q^2)$ are also connected with transverse spin sum rule \cite{Ji2012,Ji2012a,Ji2013,Leader2013,Harindranath2013,Harindranath2014} and have been verified in light-front quark-diquark model in AdS/QCD \cite{Chakrabarti2015}.

In the Drell-Yan-West frame, the charge and magnetizaton densities in the transverse plane are obtained from the two dimensional Fourier transformation of the elctromagnetic form factors (Dirac and Pauli) with respect to the momentum transfer. In a similar way, the two dimensional Fourier transform of the GFFs provides the distribution of longitudinal momentum inside the hadron. The longitudinal momentum distribution in transverse plane was first introduced in Ref. \cite{Abidin2008}, where the authors obtained the semi-empirical momentum density distribution of nucleons and for spin-1 objects in the AdS/QCD correspondence approach. A nice comparative study of charge and momentum distributions have been reported in \cite{Selyugin2009}. The longitudinal momentum densities for nucleon in the framework of soft-wall AdS/QCD has been investigated in \cite{Mondal2016a} whereas the same distribution in a light-front scalar-diquark model in AdS/QCD has been studied in \cite{Chakrabarti2015}. The longitudinal momentum distributions in transverse coordinate space have been reported in \cite{Mondal2016b}. 

In the context of $Q^2\ne0$, the 3-dimensional Fourier transformation of $J(Q^2)$ can be used to calculate the distribution of angular momentum in coordinate space \cite{Polyakov2003,Goeke2007}. But the interpretation of the 3-dimensional Fourier transformation of form factors as a distribution in 3- dimensional space becomes ambiguous and suffers from relativistic corrections for finite nucleon mass. However, a 2-dimensional Fourier transformation of $J(Q^2)$ does not suffer from such relativistic corrections. Recently, the authors in Ref.\cite{Adhikari2016} compared different definitions of the angular momentum density and concluded that none of the definitions agree at the density level. The discrepancies appeared due to the missing of total divergence terms which had been pointed out earlier in Refs.~\cite{Leader2014,LiuKeh-Fei2016}. However, in the recent work\cite{Lorce2017}, Lorc$\acute{e}$ $et.~al.$ discussed in detail different definitions of the angular momentum density and showed that the discrepancies between different definitions in the density level originate from terms that integrate to zero.

In the quark model, the nucleons consists of three quarks of two different flavors $u$ and $d$ ($p=|uud\rangle, ~n=|udd\rangle$). In the quark-diquark picture, one can schematically write, for example the proton state, $p=|d (uu)\rangle +|u (ud)\rangle$, where $(uu)$ and $(ud)$ are the diquark states.  With spin-flavor symmetry,
the diquark can be  either scalar or axial vector, and hence  both of them are required to describe the model. Since the scalar diquarks are in a
flavor singlet state and vector diquarks are in a flavor triplet state, in order to combine
to a symmetric spin-flavor wave function as demanded by the Pauli principle, the proton
state has the well-known SU(4) structure as described in \cite{Jakob1997,Bacchetta2008}. The scalar diquark alone cannot give the complete picture of a nucleon.
Recently, a phenomenological light-front quark-diquark model has been proposed in Ref. \cite{Maji2016} where both scalar and vector diquark are considered. In this model,
the light-front wavefunctions (LFWFs) for the proton are modeled from the two particle wave functions obtained in soft-wall AdS/QCD \cite{Brodsky2008a}. This
quark-diquark model is consistent with quark counting rule and Drell-Yan-West relation and it has been shown to reproduce many interesting nucleon properties \cite{Maji2016,Chakrabarti2017,Maji2017a,Mondal2017}. In last two decades, there are also numerous attempt has been made in quark-diquark models to explain the mass spectrum \cite{Jaffe:2004ph,Jaffe:2005zz,Wilczek:2004im,Selem2006,Santopinto:2004hw,Forkel:2008un,Cloet:2008reb,Anisovich:2010wxy,Ferreti:2011zy,DeSanctis:2011zz,Galata:2012xt,DeSanctis:2014ria} and the electromagnetic form factors \cite{Eichmann:2011vu,DeSanctis:2011zz}.
In the present work, we evaluate the GFFs for up and down quark in this light-front quark-diquark model considering both scalar and vector diquark. The longitudinal momentum distributions in transverse position space are studied 
in an unpolarized as well as in a transversely polarized nucleon.  We compare the results obtained in this quark-diquark model with the predictions of a soft-wall AdS/QCD model and the parametrization of GPDs based on latest global analysis by ``MRST2009" \cite{MartinA.D.2009} and the Gaussian ansatz to incorporate the $q^2$-dependence. 
We also present the spatial distribution of angular momentum for up and down quarks inside the nucleon. The different definitions of angular momentum densities are illustrated explicitly for up and down quark in this light-front quark-diquark model inspired by AdS/QCD.

The plan of the paper is as follows. A brief description of the light-front quark-diquark model has been given in section  \ref{quark_diquark_model}. We present the results of GFFs and longitudinal momentum distributions in section \ref{grav_FF} and section \ref{lmd} respectively. The illustration of different definitions angular momentum densities has been presented in section \ref{amd} and final conclusions are drawn in section \ref{conl}.
\section{Light Front Quark Diquark Model}\label{quark_diquark_model}
Here we consider a light-front quark diquark model for nucleon\cite{Maji2016} consisting with both the scalar and the vector diquark. 
Spin-0 diquark is in flavor singlet state and spin-1 diquark is in flavor triplet state. One can represent the proton state as sum of scalar-isoscalar $|u S^0 \rangle$, isoscalar-vector diquark $|u A^0\rangle$ and isovector-vector $|d A^0 \rangle$ diquark state \cite{Bacchetta2008,Jakob1997}.
The two particle Fock-state expansion for $J^z =\pm1/2$ with spin-0 diquark  is given by
\bea
|u S\rangle^\pm &=&\int \frac{dx d^2\bfp}{2(2\pi)^3\sqrt{x(1-x)}}\nonumber\\ 
&\times &\Big[ \psi^{\pm(u)}_{+}(x,\bfp)|+\frac{1}{2}~s; xP^+,\bfp\rangle \nonumber\\
&&+ \psi^{\pm(u)}_{-}(x,\bfp)|-\frac{1}{2}~s; xP^+,\bfp\rangle\Big],\label{fock_PS}
\eea
and the LF wave functions with spin-0 diquark, for $J=\pm1/2$, are given by\cite{Lepage1980}
\bea
\psi^{+(u)}_+(x,\bfp)&=& N_S~ \varphi^{(u)}_{1}(x,\bfp),\nonumber \\
\psi^{+(u)}_-(x,\bfp)&=& N_S\Big(- \frac{p^1+ip^2}{xM} \Big)\varphi^{(u)}_{2}(x,\bfp)\label{LFWF_S}\\
\psi^{-(u)}_+(x,\bfp)&=& N_S \Big(\frac{p^1-ip^2}{xM}\Big) \varphi^{(u)}_{2}(x,\bfp),\nonumber \\
\psi^{-(u)}_-(x,\bfp)&=&  N_S~ \varphi^{(u)}_{1}(x,\bfp),\nonumber
\eea
where $|\lambda_q~\lambda_S; xP^+,\bfp\rangle$ is the two particle state having struck quark of helicity $\lambda_q$ and a scalar diquark having helicity $\lambda_S=s$(spin-0 singlet diquark helicity is denoted by s to distinguish it from triplet diquark). The state with spin-1 diquark is given as \cite{Ellis2009}
\bea
|\nu~ A \rangle^\pm & = &\int \frac{dx~ d^2\bfp}{2(2\pi)^3\sqrt{x(1-x)}}\nonumber\\
&\times&\Big[ \psi^{\pm(\nu)}_{++}(x,\bfp)|+\frac{1}{2}~+1; xP^+,\bfp\rangle \nonumber\\&&+ \psi^{\pm(\nu)}_{-+}(x,\bfp)|-\frac{1}{2}~+1; xP^+,\bfp\rangle \nonumber\\
&&+\psi^{\pm(\nu)}_{+0}(x,\bfp)|+\frac{1}{2}~0; xP^+,\bfp\rangle \nonumber\\&&+ \psi^{\pm(\nu)}_{-0}(x,\bfp)|-\frac{1}{2}~0; xP^+,\bfp\rangle \nonumber\\
&&+ \psi^{\pm(\nu)}_{+-}(x,\bfp)|+\frac{1}{2}~-1; xP^+,\bfp\rangle \nonumber\\&&+ \psi^{\pm(\nu)}_{--}(x,\bfp)|-\frac{1}{2}~-1; xP^+,\bfp\rangle  \Big].\label{fock_AS}
\eea
where $|\lambda_q~\lambda_D; xP^+,\bfp\rangle$ represents a two-particle state with a quark of helicity $\lambda_q=\pm\frac{1}{2}$ and a vector diquark of helicity $\lambda_D=\pm 1,0$.
The LFWFs for $J=+1/2$ are
\bea 
\psi^{+(\nu)}_{+~+}(x,\bfp)&=& N^{(\nu)}_1 \sqrt{\frac{2}{3}} \Big(\frac{p^1-ip^2}{xM}\Big) \varphi^{(\nu)}_{2}(x,\bfp),\nonumber \\
\psi^{+(\nu)}_{-~+}(x,\bfp)&=& N^{(\nu)}_1 \sqrt{\frac{2}{3}} \varphi^{(\nu)}_{1}(x,\bfp),\nonumber \\
\psi^{+(\nu)}_{+~0}(x,\bfp)&=& - N^{(\nu)}_0 \sqrt{\frac{1}{3}} \varphi^{(\nu)}_{1}(x,\bfp),\label{LFWF_Vp}\\
\psi^{+(\nu)}_{-~0}(x,\bfp)&=& N^{(\nu)}_0 \sqrt{\frac{1}{3}} \Big(\frac{p^1+ip^2}{xM} \Big)\varphi^{(\nu)}_{2}(x,\bfp),\nonumber \\
\psi^{+(\nu)}_{+~-}(x,\bfp)&=& 0,\nonumber \\
\psi^{+(\nu)}_{-~-}(x,\bfp)&=&  0, \nonumber 
\eea
and for $J=-1/2$ are
\bea 
\psi^{-(\nu)}_{+~+}(x,\bfp)&=& 0,\nonumber \\
\psi^{-(\nu)}_{-~+}(x,\bfp)&=& 0,\nonumber \\
\psi^{-(\nu)}_{+~0}(x,\bfp)&=& N^{(\nu)}_0 \sqrt{\frac{1}{3}} \Big( \frac{p^1-ip^2}{xM} \Big) \varphi^{(\nu)}_{2}(x,\bfp),\label{LFWF_Vm}\\
\psi^{-(\nu)}_{-~0}(x,\bfp)&=& N^{(\nu)}_0\sqrt{\frac{1}{3}} \varphi^{(\nu)}_{1}(x,\bfp),\nonumber \\
\psi^{-(\nu)}_{+~-}(x,\bfp)&=& - N^{(\nu)}_1 \sqrt{\frac{2}{3}} \varphi^{(\nu)}_{1}(x,\bfp),\nonumber \\
\psi^{-(\nu)}_{-~-}(x,\bfp)&=& N^{(\nu)}_1 \sqrt{\frac{2}{3}} \Big(\frac{p^1+ip^2}{xM}\Big) \varphi^{(\nu)}_{2}(x,\bfp),\nonumber
\eea
having flavor index $\nu=u,d$.
The LFWFs $\varphi^{(\nu)}_i(x,\bfp)$ are  a modified form of the  soft-wall AdS/QCD prediction\cite{Gutsche2014,Gutsche2015}
\bea
\varphi_i^{(\nu)}(x,\bfp)&=&\frac{4\pi}{\kappa}\sqrt{\frac{\log(1/x)}{1-x}}x^{a_i^\nu}(1-x)^{b_i^\nu}\nonumber\\&\times&\exp\Big[-\delta^\nu\frac{\bfp^2}{2\kappa^2}\frac{\log(1/x)}{(1-x)^2}\Big].
\label{LFWF_phi}
\eea
The wave functions $\varphi_i^\nu ~(i=1,2)$ reduce to the AdS/QCD prediction\cite{Brodsky2008a} for the parameters $a_i^\nu=b_i^\nu=0$  and $\delta^\nu=1.0$.
 We use the AdS/QCD scale parameter $\kappa =0.4~GeV$ as determined in \cite{Chakrabarti2013} and the quarks are  assumed  to be  massless.

 \section{Gravitational form factors}\label{grav_FF}
 The matrix elements of local operators like energy momentum tensor, electromagnetic current and moment of structure functions have exact representation in light-front Fock state wave functions of bound states such as hadrons. One can obtain the GFFs by calculating the matrix element of the energy-momentum tensor. 
The second moment of spin non-flip and spin flip GPDs also gives the gravitational form factors $A(Q^2)$ and $B(Q^2)$ respectively. These form factors are part of the energy-momentum tensor \cite{Brodsky2001}
\bea
&&\langle P' | T^{\mu \nu} |P \rangle = \bar{U}(P')\Bigg[A(q^2) \bar{P}^\mu \gamma^\nu+ B(q^2) \frac{i \sigma^{\mu \alpha} q_\alpha P^\nu}{2 M}\nonumber\\&&+ C(q^2) (q^\mu q^\nu- g^{\mu \nu} q^2)+\bar{C}(q^2) M g^{\mu \nu}\Bigg] U(P),
\label{gff}
\eea
where, $\bar{P}^\mu=\frac{1}{2}(P'+P)^\mu$, $q^\mu=(P'-P)^\mu$, $a ^{(\mu}b^{\nu)}=\frac{1}{2}(a^\mu b^\nu+ a^\nu b^\mu)$ and $U(P)$ is the spinor. By calculating the $(++)$ component of energy-momentum tensor, one can obtain
\bea
\langle P+q, \uparrow|\frac{T_i^{++}(0)}{2 (P^+)^2}|P, \uparrow \rangle &=& A_i(q^2), \nonumber\\
\langle P+q, \uparrow|\frac{T_i^{++}(0)}{2 (P^+)^2}|P, \downarrow \rangle &=& -(q^1 - i q^2)\frac{B_i(q^2)}{2M}.
\label{gff1}
\eea
The $A_i(q^2)$ and $B_i(q^2)$ in Eq. (\ref{gff1}) are the form factors which are very similar to the Dirac and Pauli form factors. The Dirac and Pauli form factors can be obtained from the helicity non-flip and helicity flip vector current matrix elements of the $J^+$ current. 
Using the two particle Fock states in Eqs.(\ref{fock_PS}) and (\ref{fock_AS}) we evaluate the flavor contributions to GFFs $A(q^2)$ and $B(q^2)$ in terms of the overlap of the wavefunctions for scalar diquark as \cite{Brodsky2001}
\bea 
&&A_q^{u(S)}(Q^2)=\int\frac{d^2\bfp dx}{16\pi^3}~x\Big[\psi^{+(u)\dagger}_+(x,\bfp^\prime)\psi^{+(u)}_+(x,\bfp)\nonumber\\&&+\psi^{+(u)\dagger}_-(x,\bfp^\prime)\psi^{+(u)}_-(x,\bfp)\Big],\\
&&\nonumber\\
&&B_q^{u(S)}(Q^2)=-\frac{2M}{q^1-iq^2}\int\frac{d^2\bfp dx}{16\pi^3}~x\Big[\psi^{+(u)\dagger}_+(x,\bfp^\prime)\nonumber\\&&\psi^{-(u)}_+(x,\bfp)+\psi^{+(u)\dagger}_-(x,\bfp^\prime)\psi^{-(u)}_-(x,\bfp)\Big],
\eea
and for vector diquark
\bea 
&&A_q^{\nu(A)}(Q^2)=\int\frac{d^2\bfp dx}{16\pi^3}~x\Big[\psi^{+(\nu)\dagger}_{++}(x,\bfp^\prime)\psi^{+(\nu)}_{++}(x,\bfp)
+\nonumber\\&&\psi^{+(\nu)\dagger}_{-+}(x,\bfp^\prime)\psi^{+(\nu)}_{-+}(x,\bfp)+\psi^{+(\nu)\dagger}_{+0}(x,\bfp^\prime)\psi^{+(\nu)}_{+0}(x,\bfp)\nonumber\\&&+\psi^{+(\nu)\dagger}_{-0}(x,\bfp^\prime)\psi^{+(\nu)}_{-0}(x,\bfp)\Big],\\
&&\nonumber\\
&&B_q^{\nu(A)}(Q^2)=-\frac{2M}{q^1-iq^2}\int\frac{d^2\bfp dx}{16\pi^3}~x\Big[\psi^{+(\nu)\dagger}_{+0}(x,\bfp^\prime) \nonumber\\&&\psi^{-(\nu)}_{+0}(x,\bfp)+\psi^{+(\nu)\dagger}_{-0}(x,\bfp^\prime)\psi^{-(\nu)}_{-0}(x,\bfp)\Big],
\eea
where $\bfp^\prime = \bfp+(1-x)\bfq$. Similarly, the diquark contributions to GFFs $A(q^2)$ and $B(q^2)$ can be written in terms of LFWFs for scalar diquark as
\bea 
&&A_{qq}^{u(S)}(Q^2)=\int\frac{d^2\bfp dx}{16\pi^3}~(1-x)\Big[\psi^{+(u)\dagger}_+(x,\bfp^{\prime\prime})\nonumber\\&&\psi^{+(u)}_+(x,\bfp)+\psi^{+(u)\dagger}_-(x,\bfp^{\prime\prime})\psi^{+(u)}_-(x,\bfp)\Big],\\
&&\nonumber\\
&&B_{qq}^{u(S)}(Q^2)=-\frac{2M}{q^1-iq^2}\int\frac{d^2\bfp dx}{16\pi^3}~(1-x)\Big[\psi^{+(u)\dagger}_+(x,\bfp^{\prime\prime})\nonumber\\
&&\psi^{-(u)}_+(x,\bfp)+\psi^{+(u)\dagger}_-(x,\bfp^{\prime\prime})\psi^{-(u)}_-(x,\bfp)\Big],
\eea
and for vector diquark
\bea 
&&A_{qq}^{\nu(A)}(Q^2)=\int\frac{d^2\bfp dx}{16\pi^3}~(1-x)\Big[\psi^{+(\nu)\dagger}_{++}(x,\bfp^{\prime\prime})\nonumber\\&&
\psi^{+(\nu)}_{++}(x,\bfp)+\psi^{+(\nu)\dagger}_{-+}(x,\bfp^{\prime\prime})\psi^{+(\nu)}_{-+}(x,\bfp)+\nonumber\\
&&\psi^{+(\nu)\dagger}_{+0}(x,\bfp^{\prime\prime})\psi^{+(\nu)}_{+0}(x,\bfp)+\psi^{+(\nu)\dagger}_{-0}(x,\bfp^{\prime\prime})\psi^{+(\nu)}_{-0}(x,\bfp)\Big],\nonumber\\
&&\\
&&B_{qq}^{\nu(A)}(Q^2)=-\frac{2M}{q^1-iq^2}\int\frac{d^2\bfp dx}{16\pi^3}~(1-x)\Big[\psi^{+(\nu)\dagger}_{+0}(x,\bfp^{\prime\prime})\nonumber\\&&\psi^{-(\nu)}_{+0}(x,\bfp)+\psi^{+(\nu)\dagger}_{-0}(x,\bfp^{\prime\prime})\psi^{-(\nu)}_{-0}(x,\bfp)\Big],
\eea
where $\bfp^{\prime\prime} = \bfp-x\bfq$ and $Q^2=\bfq^2=-t$.
The superscripts  $A=V,VV$ for isoscalar-vector diquark and isovector-vector diquark respectively. 
 In the SU(4) structure, depending on different flavors (struck quark) quark and diquark GFFs are written in terms of scalar and vector diquarks as\cite{Bacchetta2008}
\bea 
A^f_{u}(Q^2)&=&C^2_S A_q^{u(S)}(Q^2) + C^2_V A_q^{u(V)}(Q^2),\label{Afu}\\
A^f_{d}(Q^2)&=&C^2_{VV} A_q^{d(VV)}(Q^2).\label{Afd}\\
&&\nonumber\\
A^f_{ud}(Q^2)&=&C^2_S A_{qq}^{u(S)}(Q^2) + C^2_V A_{qq}^{u(V)}(Q^2),\label{Afud}\\
A^f_{uu}(Q^2)&=&C^2_{VV} A_{qq}^{d(VV)}(Q^2).\label{Afuu}
\eea
$B(Q^2)$ also follows the same expressions as in Eqs.(\ref{Afu}-\ref{Afuu}). Using the LFWFs given in Eqs.(\ref{LFWF_S},\ref{LFWF_Vp},\ref{LFWF_Vm}), 
the explicit calculation gives
\bea 
A_q^{u(S)}(Q^2)&=&N^2_S  R_1^{(u)}(Q^2),
\nonumber\\
B_q^{u(S)}(Q^2)&=& N^2_S R_2^{(u)}(Q^2),
\nonumber\\
A_q^{u(V)}(Q^2)&=& (\frac{1}{3} N^{(u)2}_0+ \frac{2}{3} N^{(u)2}_1) R_1^{(u)}(Q^2),
\nonumber\\ 
B_q^{u(V)}(Q^2)&=& - \frac{1}{3} N^{(u)2}_0 R_2^{(u)}(Q^2),
\\
A_q^{d(VV)}(Q^2)&=&(\frac{1}{3} N^{(d)2}_0+ \frac{2}{3} N^{(d)2}_1) R_1^{(d)}(Q^2), 
\nonumber\\
B_q^{d(VV)}(Q^2)&=& - \frac{1}{3} N^{(d)2}_0 R_2^{(d)}(Q^2), 
\nonumber\\
&&\nonumber\\
A_{qq}^{u(S)}(Q^2)&=&N^2_S  R_3^{(u)}(Q^2)\nonumber\\
B_{qq}^{u(S)}(Q^2)&=&-N^2_S R_4^{(u)}(Q^2),\nonumber\\
A_{qq}^{u(V)}(Q^2)&=& (\frac{1}{3} N^{(u)2}_0+ \frac{2}{3} N^{(u)2}_1) R_3^{(u)}(Q^2),\nonumber\\
B_{qq}^{u(V)}(Q^2)&=& \frac{1}{3} N^{(u)2}_0 R_4^{(u)}(Q^2),\\
A_{qq}^{d(VV)}(Q^2)&=&(\frac{1}{3} N^{(d)2}_0+ \frac{2}{3} N^{(d)2}_1) R_3^{(d)}(Q^2),\nonumber\\
B_{qq}^{d(VV)}(Q^2)&=& \frac{1}{3} N^{(d)2}_0 R_4^{(d)}(Q^2),\nonumber
\eea
where superscript $S,V$ and $VV$ represent the contributions with isoscalar-scalar diquark, isoscalar-vector diquark and isovector-vector diquarks respectively and $R_i^{(\nu)}(Q^2)$ are given by 
\bea 
&& R_1^{(\nu)}(Q^2)=\int dx \Big[x^{2a^{\nu}_1+1}(1-x)^{2b^{\nu}_1+1}\frac{1}{\delta^\nu}\nonumber\\&&+x^{2a^{\nu}_2-1}(1-x)^{2b^{\nu}_2+3} \frac{\kappa^2}{(\delta^\nu)^2 M^2\log(1/x)} \nonumber\\
&&\times \Big(1-\delta^\nu\frac{Q^2}{4\kappa^2} \log(1/x)\Big)\Big]\exp\Big[-\delta^\nu\frac{Q^2}{4\kappa^2} \log(1/x)\Big],\label{R1}\\
&&\nonumber\\
&& R_2^{(\nu)}(Q^2)=2\int dx~ x^{a^{\nu}_1+a^{\nu}_2}(1-x)^{b^{\nu}_1+b^{\nu}_2+2}\frac{1}{\delta^\nu}\nonumber\\&&\times \exp\Big[-\delta^\nu\frac{Q^2}{4\kappa^2} \log(1/x)\Big],\\
&&\nonumber\\
&& R_3^{(\nu)}(Q^2)=\int dx \Big[x^{2a^{\nu}_1}(1-x)^{2b^{\nu}_1+2}\frac{1}{\delta^\nu}\nonumber\\&&+x^{2a^{\nu}_2-2}(1-x)^{2b^{\nu}_2+4} \frac{\kappa^2}{(\delta^\nu)^2 M^2\log(1/x)} \nonumber\\
&&\times \Big(1-\delta^\nu\frac{x^2 Q^2}{4\kappa^2(1-x)^2} \log(1/x)\Big)\Big]\nonumber\\&&\times\exp\Big[-\delta^\nu\frac{x^2Q^2}{4\kappa^2(1-x)^2} \log(1/x)\Big],\label{R3}\\
&&\nonumber\\
&& R_4^{(\nu)}(Q^2)=2\int dx~ x^{a^{\nu}_1+a^{\nu}_2}(1-x)^{b^{\nu}_1+b^{\nu}_2+2}\frac{1}{\delta^\nu}\nonumber\\&&\times \exp\Big[-\delta^\nu\frac{x^2Q^2}{4\kappa^2(1-x)^2} \log(1/x)\Big]. 
\eea
The values of parameters $a_i$, $b_i$, $N_s, N_0^\nu, N_1^\nu$ and coefficients $C_i^2$ are obtained from Ref. \cite{Maji2016}. The total GFFs  for proton for the
struck $u$ and $d$ quarks are given as
\bea
A_{tot}^u(Q^2)&=&\frac{1}{2} [A_{u}^f+A_{ud}^f]=A_{u}+A_{ud},\nonumber\\
A_{tot}^d(Q^2)&=& [A_{d}^f+A_{uu}^f]=A_{d}+A_{uu}.\label{tot}
\eea
For $A_{tot}^u$, the $\frac{1}{2}$ factor appears due to two possibilities of the struck quark being a $u$ quark
(two valence $u$ quark in proton). $A_{q}^f$ and $A_{q}$ represent each flavor and each quark GFF respectively and $A_{u}^f=2A_{u}$, $A_{d}^f=A_{d}$. Similarly for diquark, $A_{ud}^f=2A_{ud}$ and $A_{uu}^f=A_{uu}$. One can also write a similar expression for GFF $B(Q^2)$ as $A(Q^2)$ in Eq.(\ref{tot}).
\begin{figure*}[htbp]
\begin{minipage}[c]{0.98\textwidth}
{(a)}\includegraphics[width=7.5cm,clip]{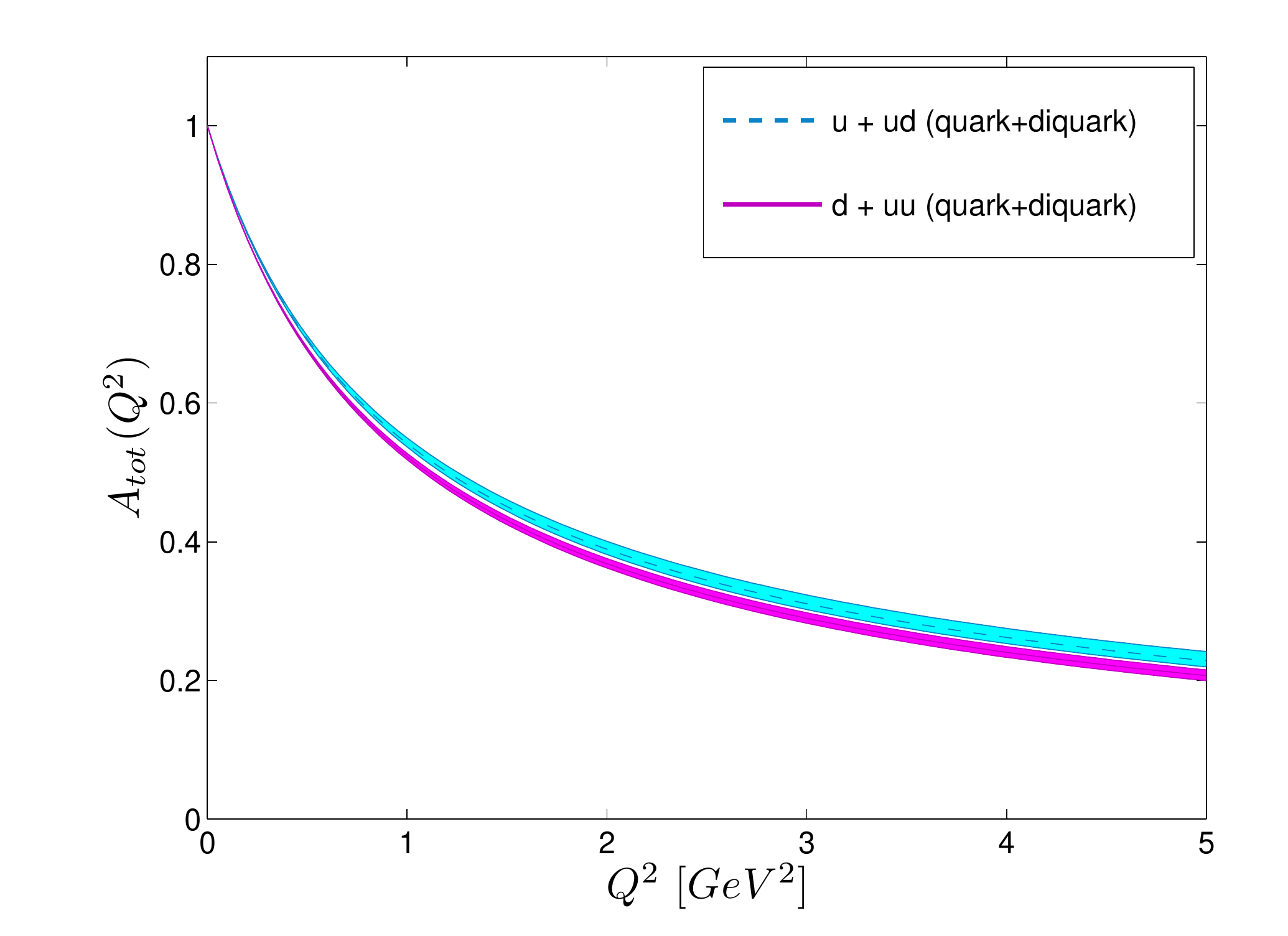}
{(b)}\includegraphics[width=7.5cm,clip]{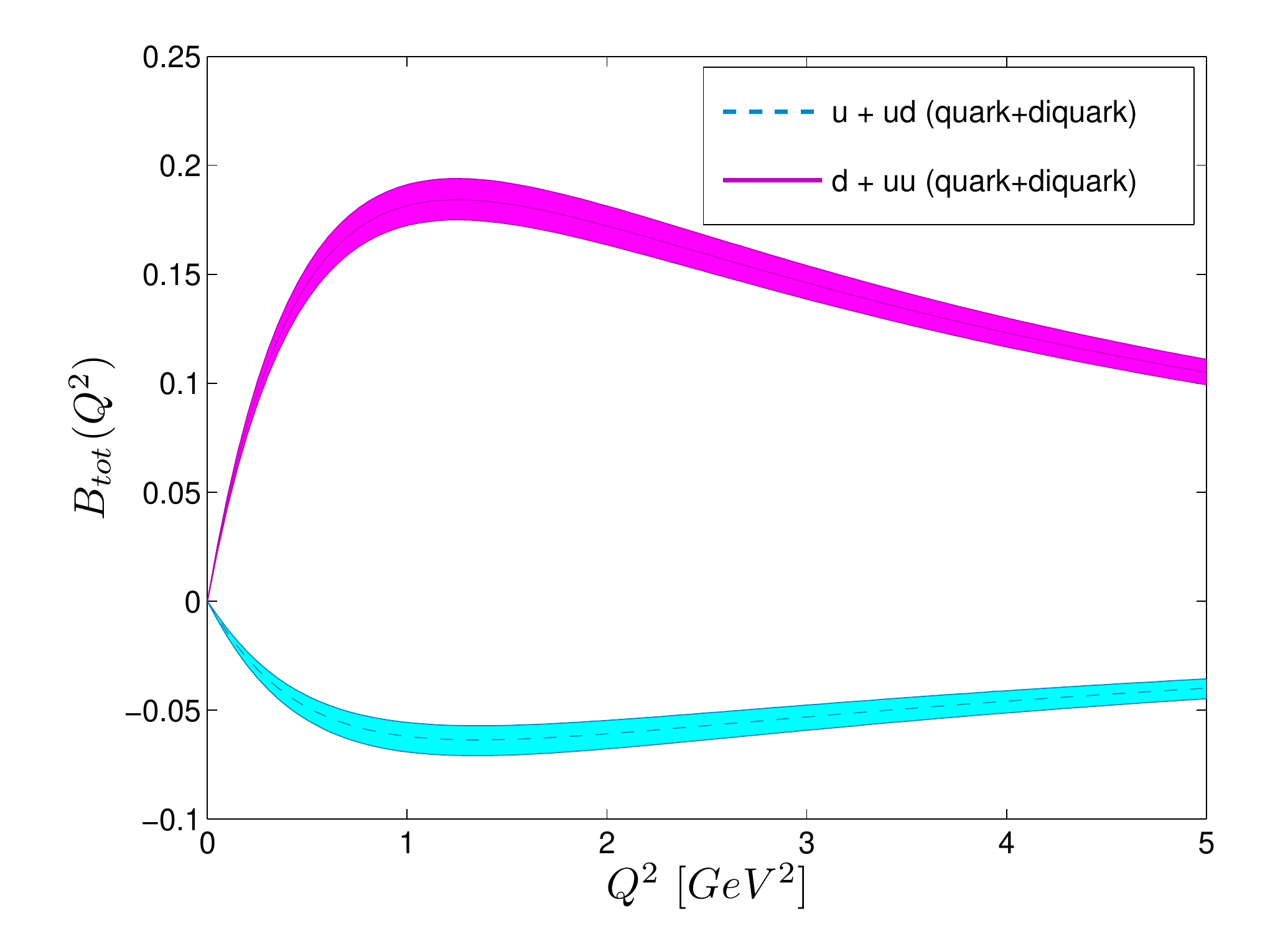}
\end{minipage}
\begin{minipage}[c]{0.98\textwidth}
{(c)}\includegraphics[width=7.5cm,clip]{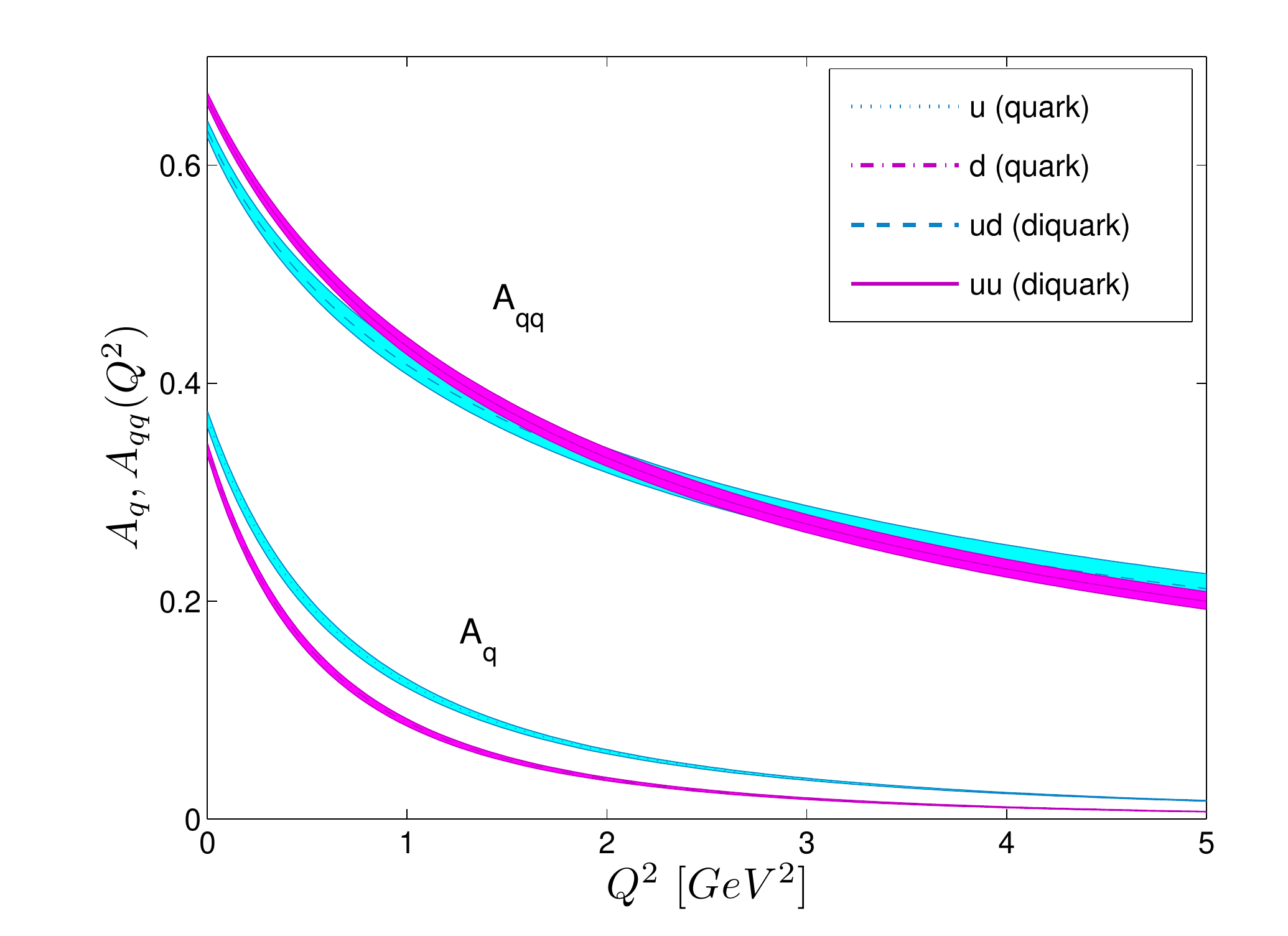}
{(d)}\includegraphics[width=7.5cm,clip]{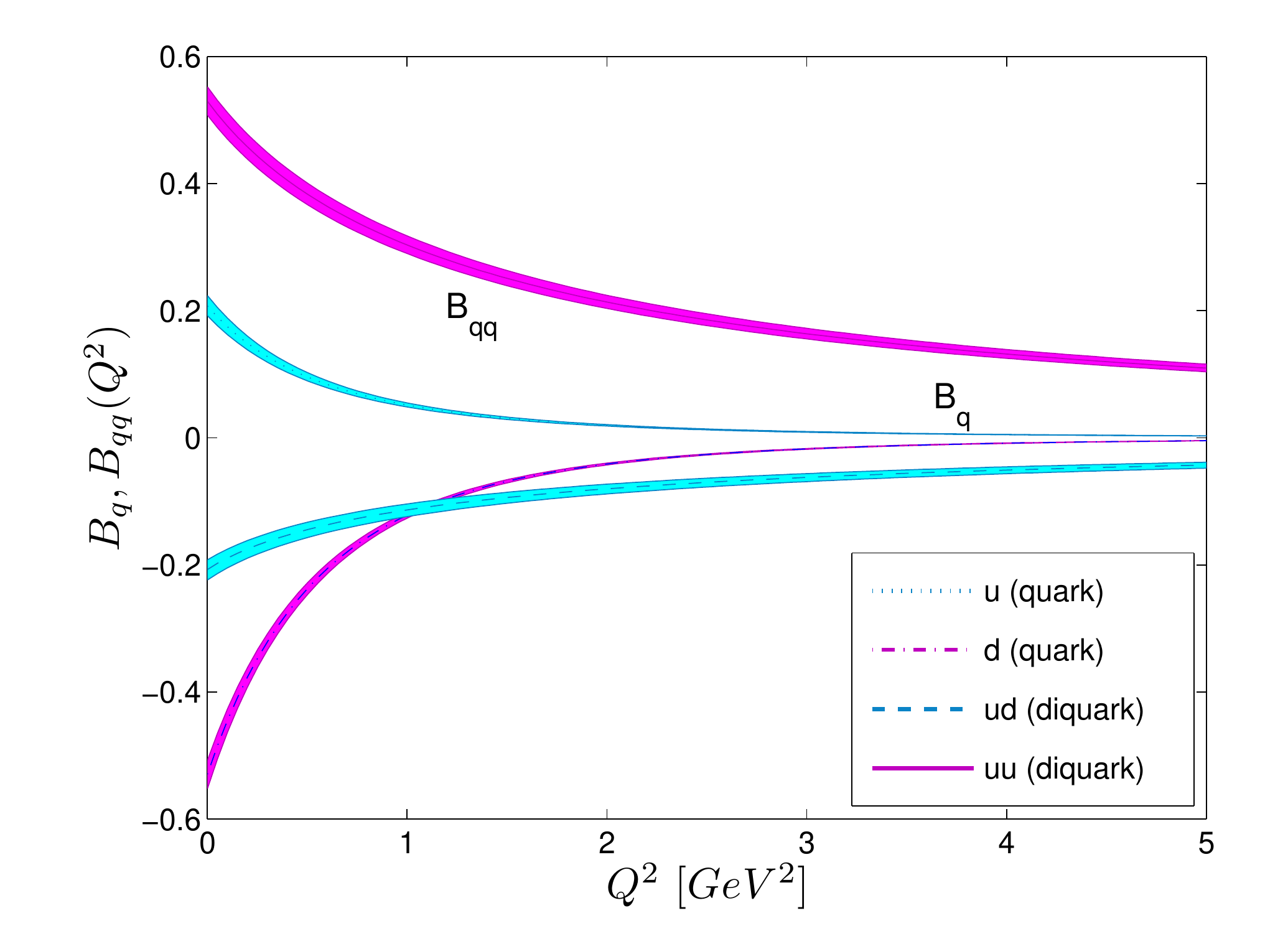}
\end{minipage}
\caption{\label{GFFs}Plots of gravitational form factors, (a) $A_{tot}(Q^2)$, (b) $B_{tot}(Q^2)$, (c) and (d) represent the quark and diquark contributions to the total GFFs $A_{tot}$ and $B_{tot}$ respectively. The error bands correspond to 2$\sigma$ error in the model parameters.}
\end{figure*}
\begin{figure*}[htbp]
\begin{minipage}[c]{0.98\textwidth}
{(a)}\includegraphics[width=7.5cm,clip]{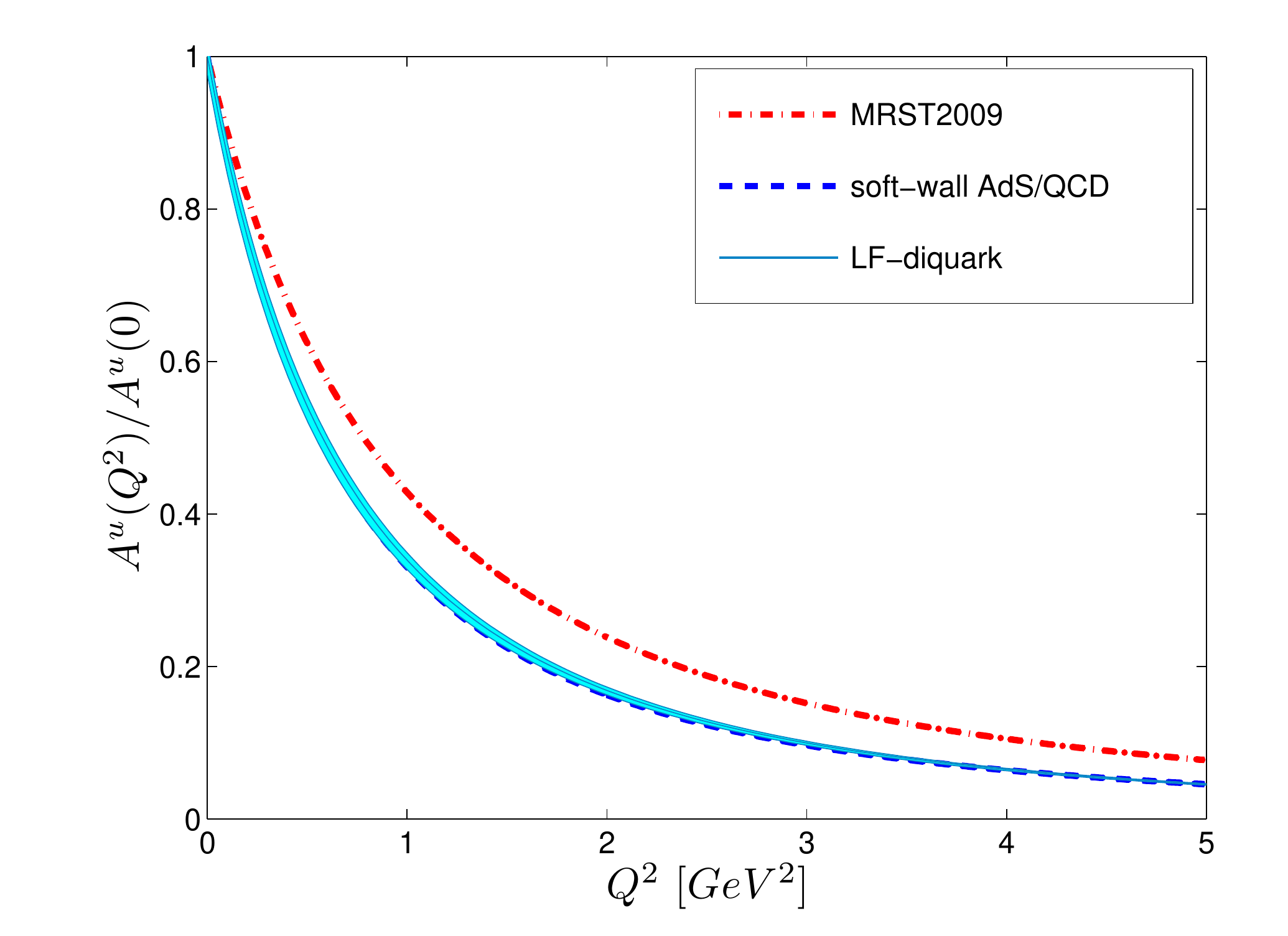}
{(b)}\includegraphics[width=7.5cm,clip]{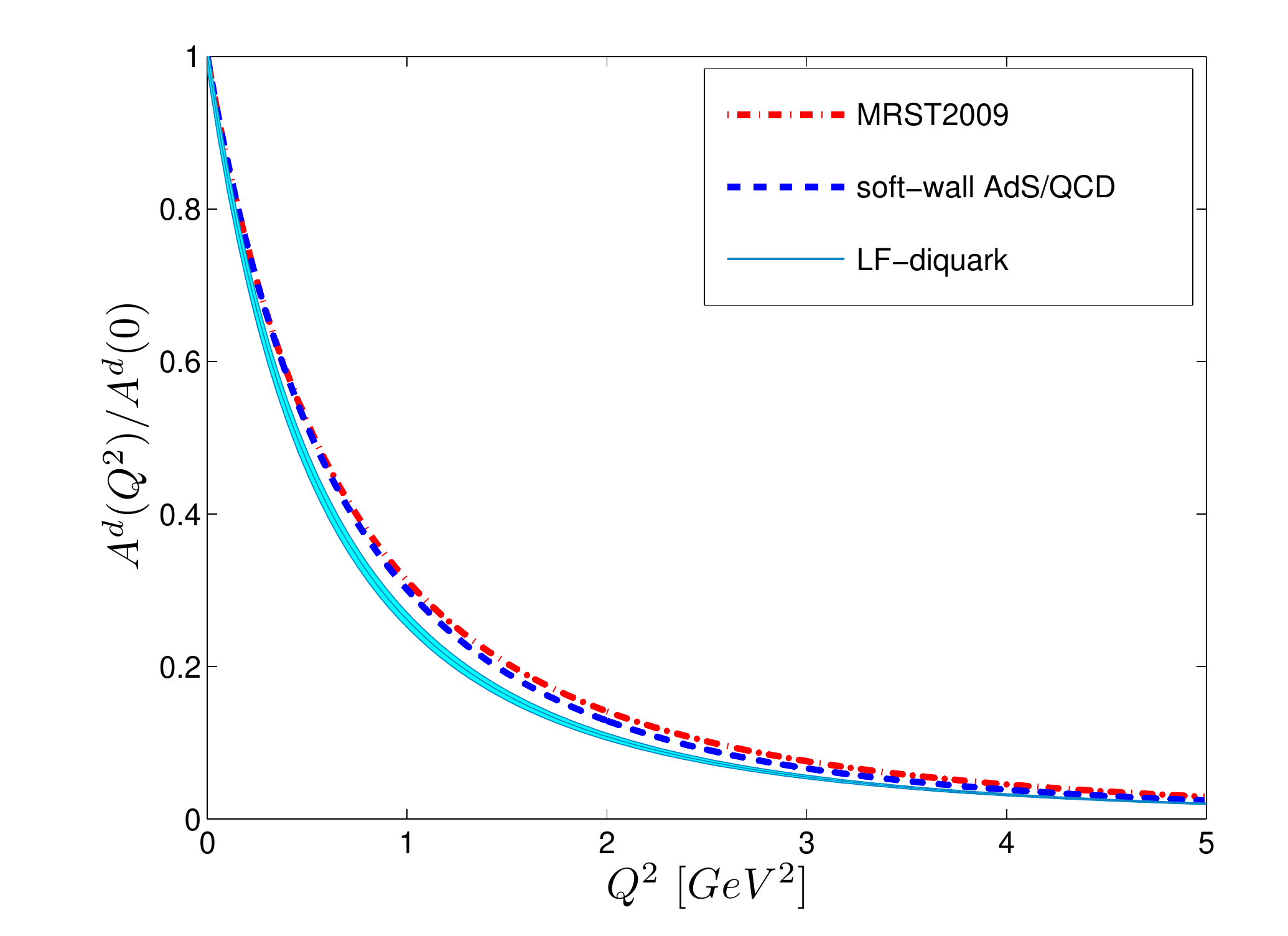}
\end{minipage}
\begin{minipage}[c]{0.98\textwidth}
{(c)}\includegraphics[width=7.5cm,clip]{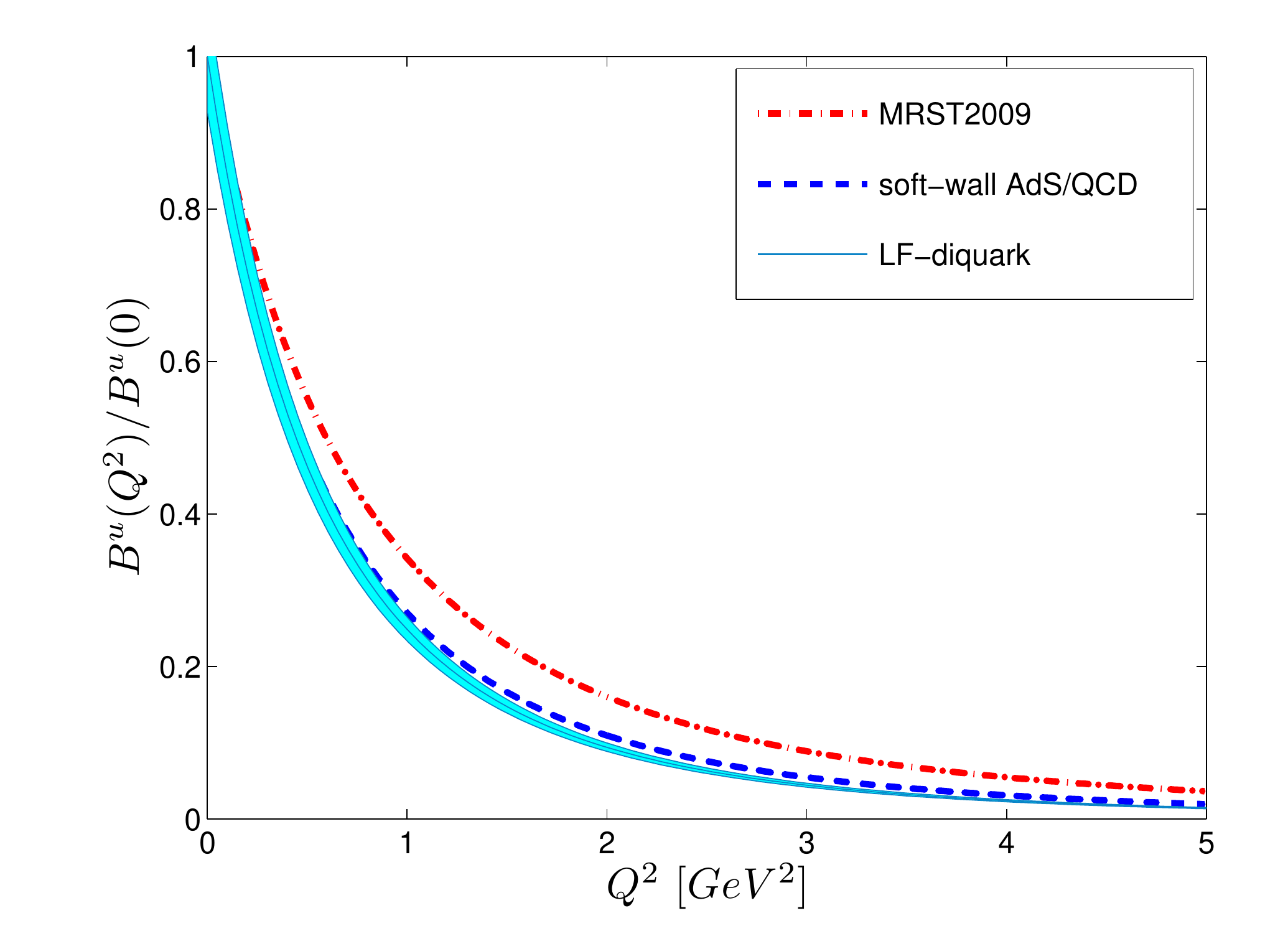}
{(d)}\includegraphics[width=7.5cm,clip]{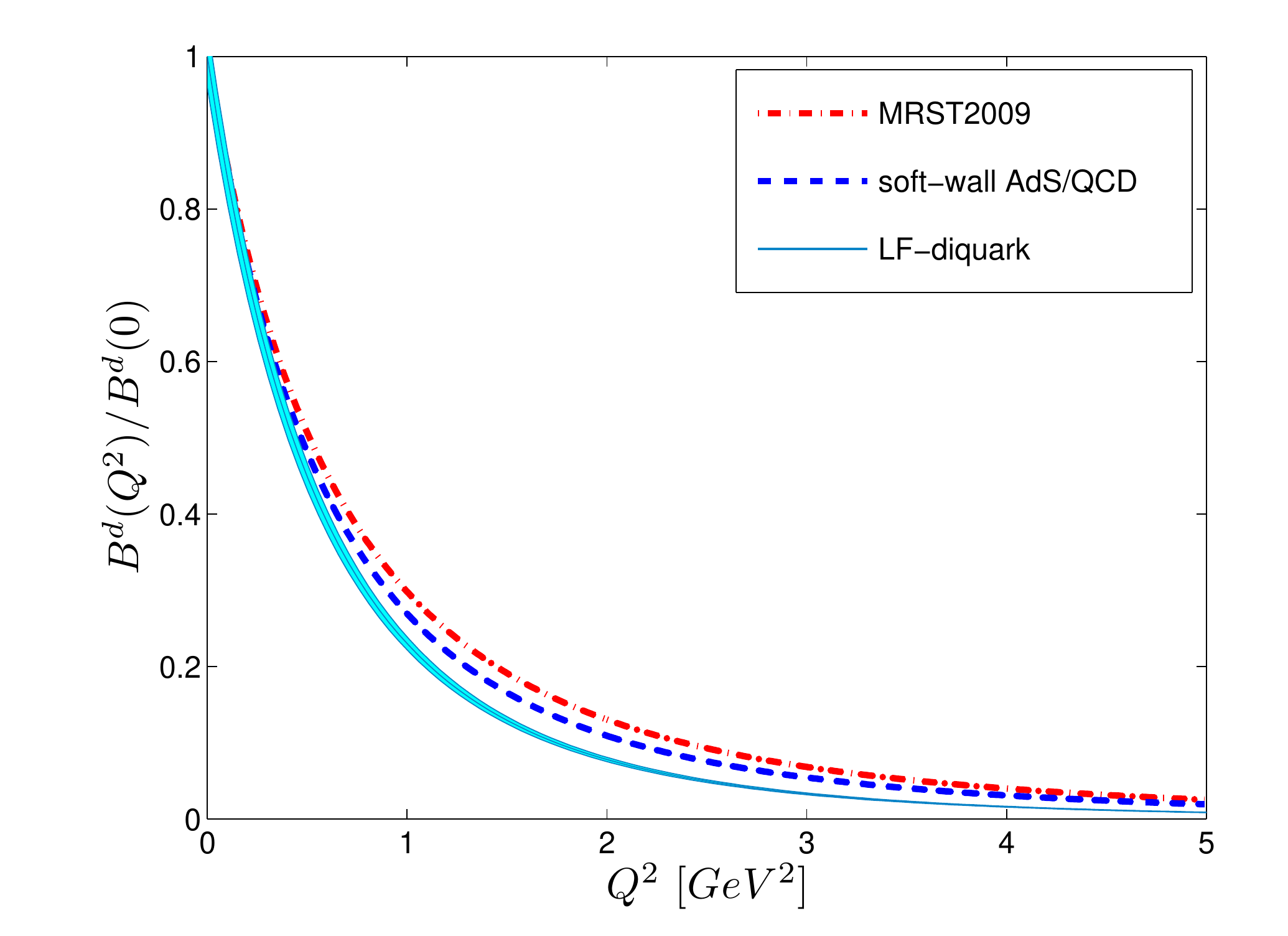}
\end{minipage}
\caption{\label{GFFs_AB}Comparison of the gravitational form factors for up and down quarks. The blue solid line is for quark-diquark model. The red dot-dashed and blue dashed lines represent the GPDs parametrization based on ``MRST2009'' \cite{neetika2016} and the soft-wall AdS/QCD model \cite{Sharma2014}. The error bands correspond to 2$\sigma$ error in the quark-diquark model parameters.}
\end{figure*}
\begin{figure*}[htbp]
\begin{minipage}[c]{0.98\textwidth}
{(a)}\includegraphics[width=7.5cm,clip]{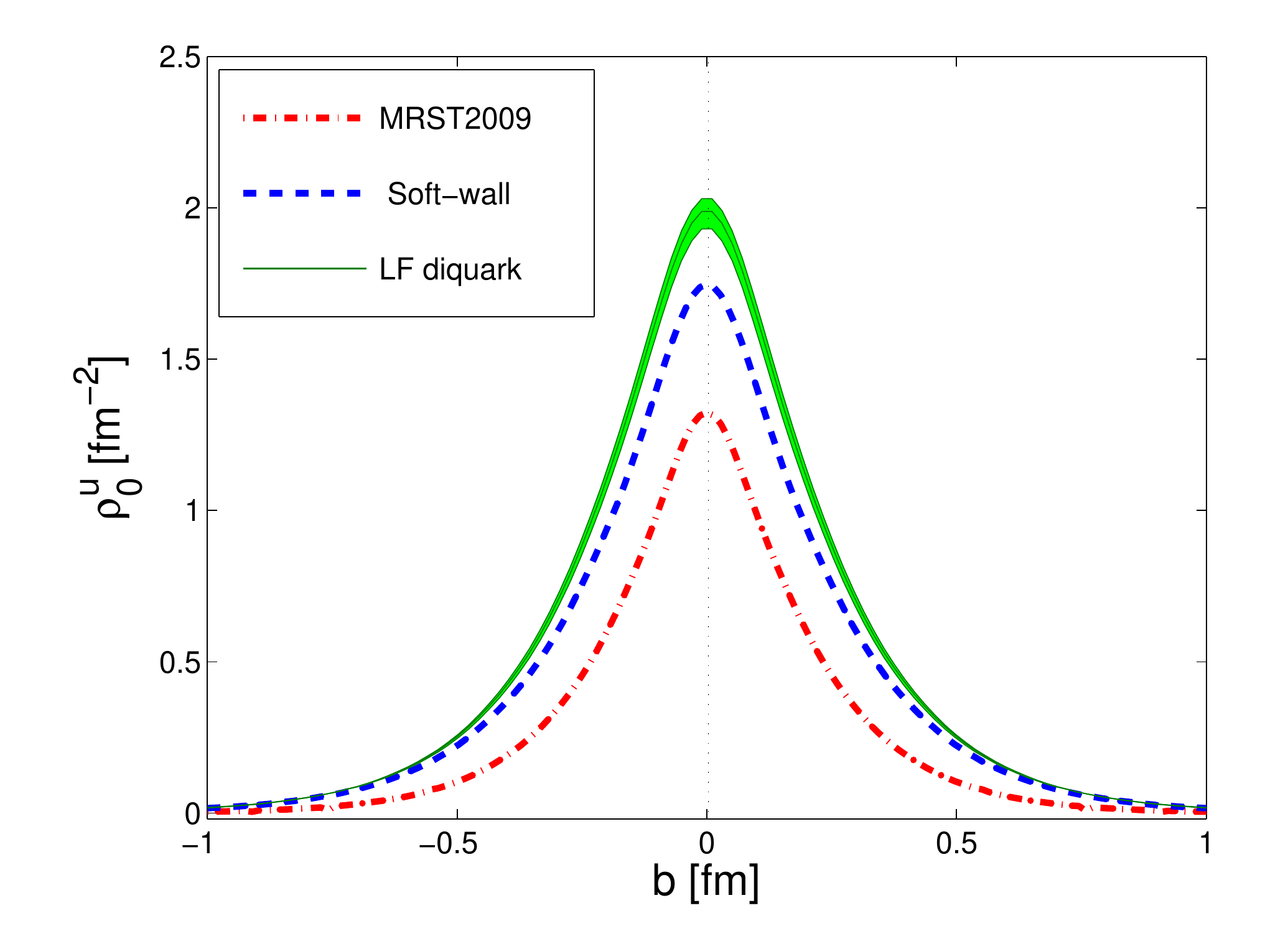}
{(b)}\includegraphics[width=7.5cm,clip]{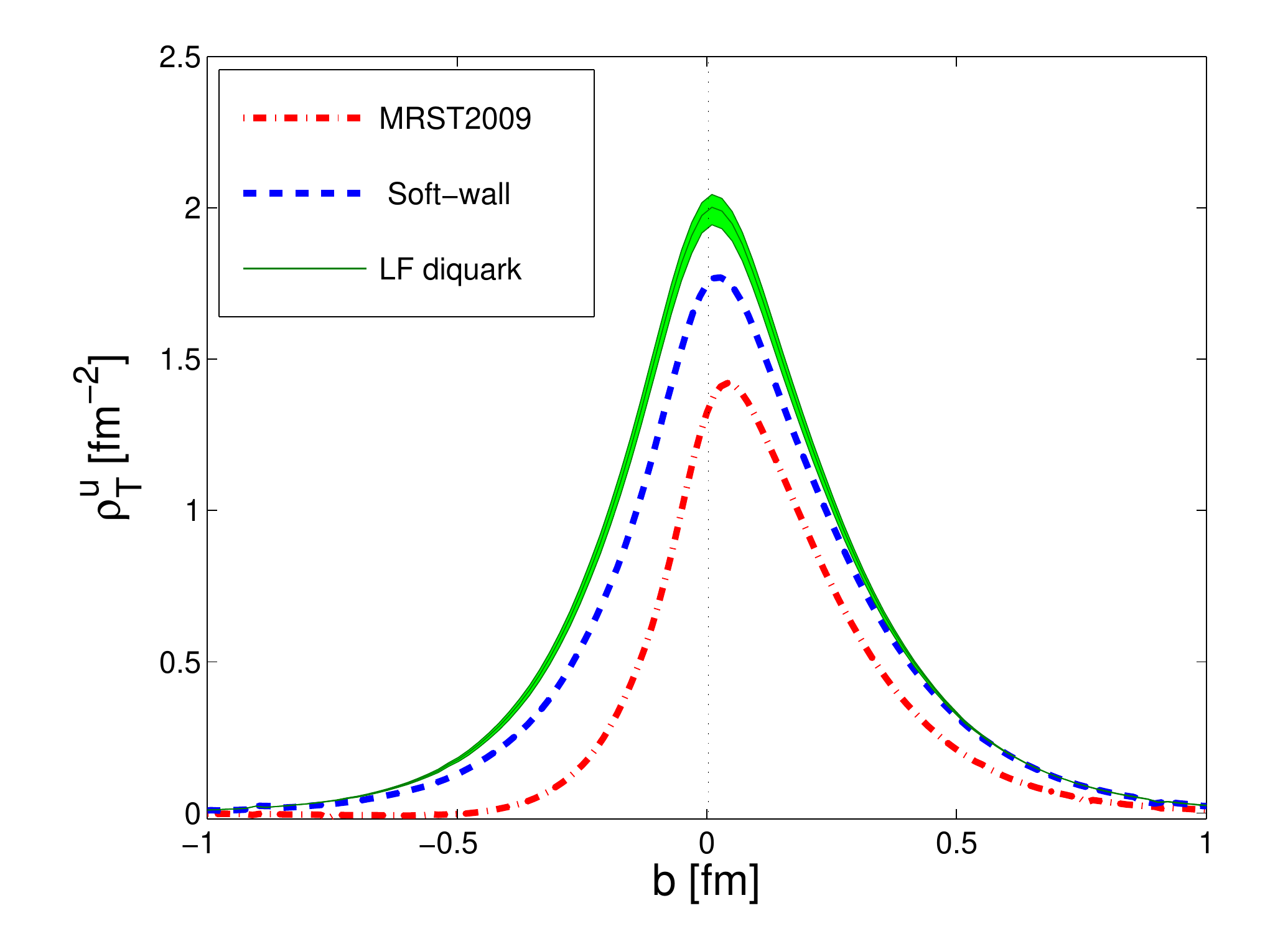}
\end{minipage}
\begin{minipage}[c]{0.98\textwidth}
{(c)}\includegraphics[width=7.5cm,clip]{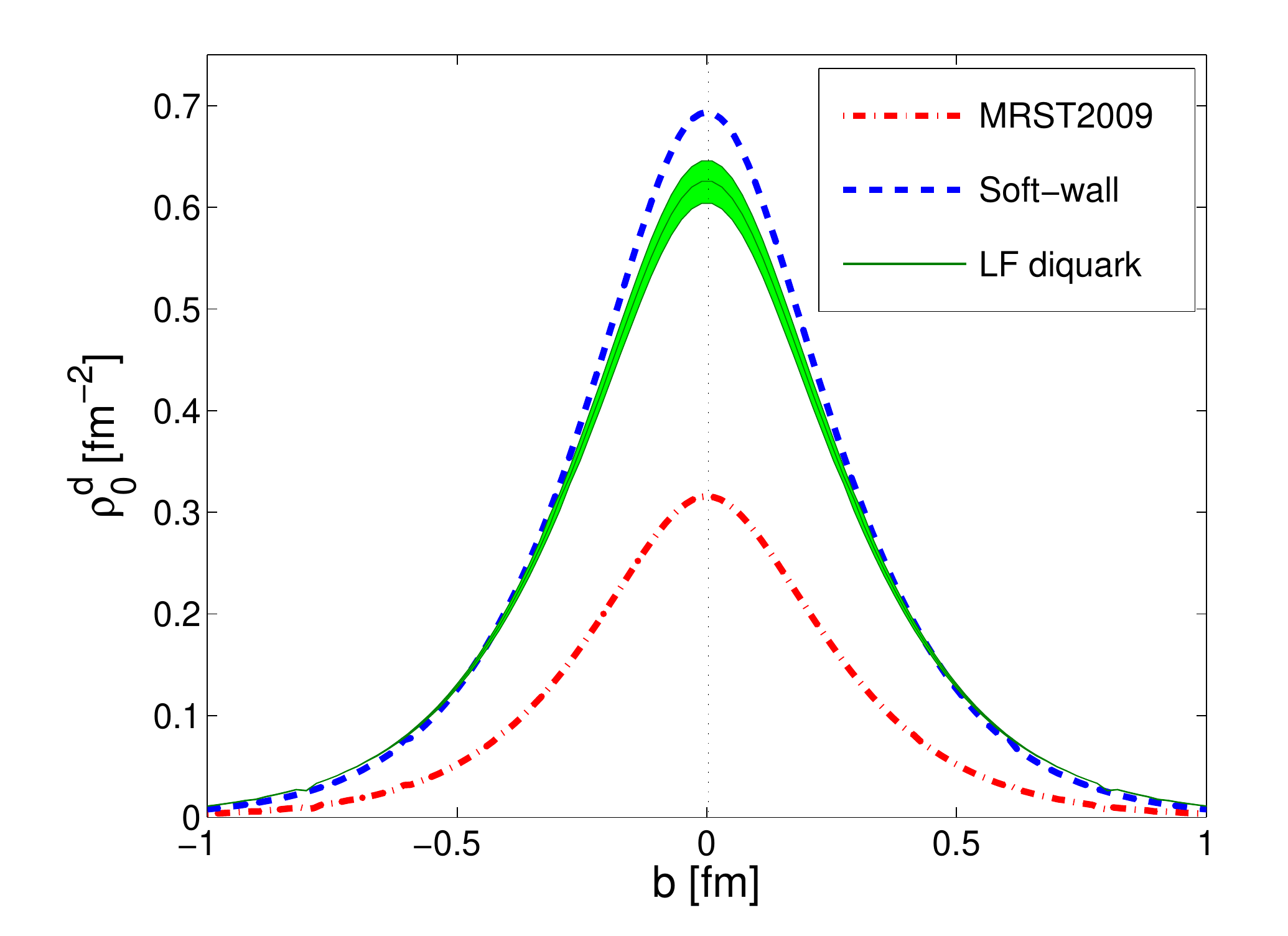}
{(d)}\includegraphics[width=7.5cm,clip]{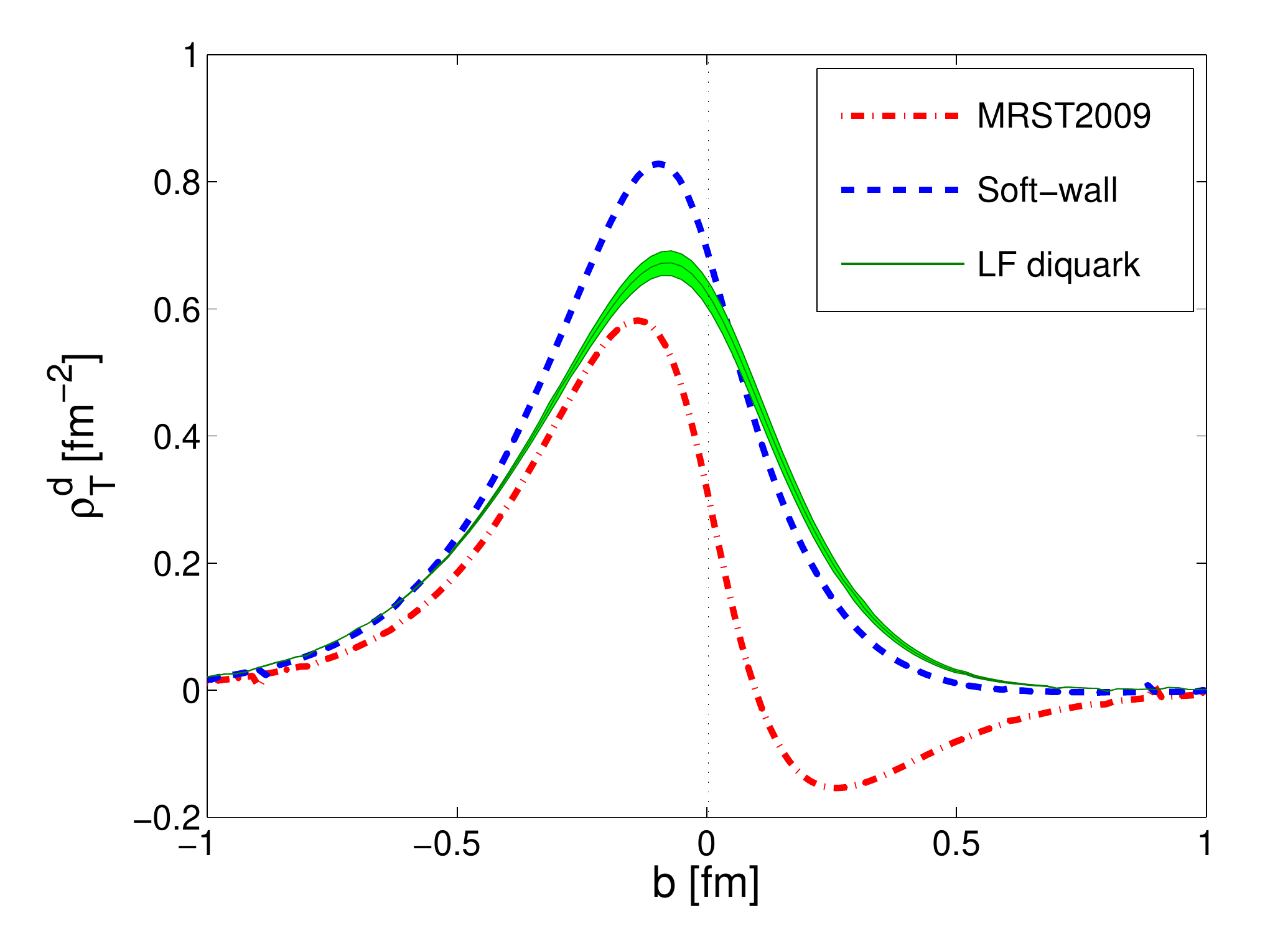}
\end{minipage}

\caption{\label{LMD}Plots of the longitudinal momentum densities for up quark (upper panel) and down quark (lower panel) in an unpolarized (left panel) and a transversely polarized (right panel) nucleon polarized along $\hat{x}$-direction calculated in LF-diquark model(solid line), soft-wall AdS/QCD model(dashed line) and GPDs parametrizations(dashed dot line). The error bands correspond to 2$\sigma$ error in the quark-diquark model parameters.}
\end{figure*}
\begin{figure*}[htbp]
\begin{minipage}[c]{0.98\textwidth}
{(a)}\includegraphics[width=7.5cm,clip]{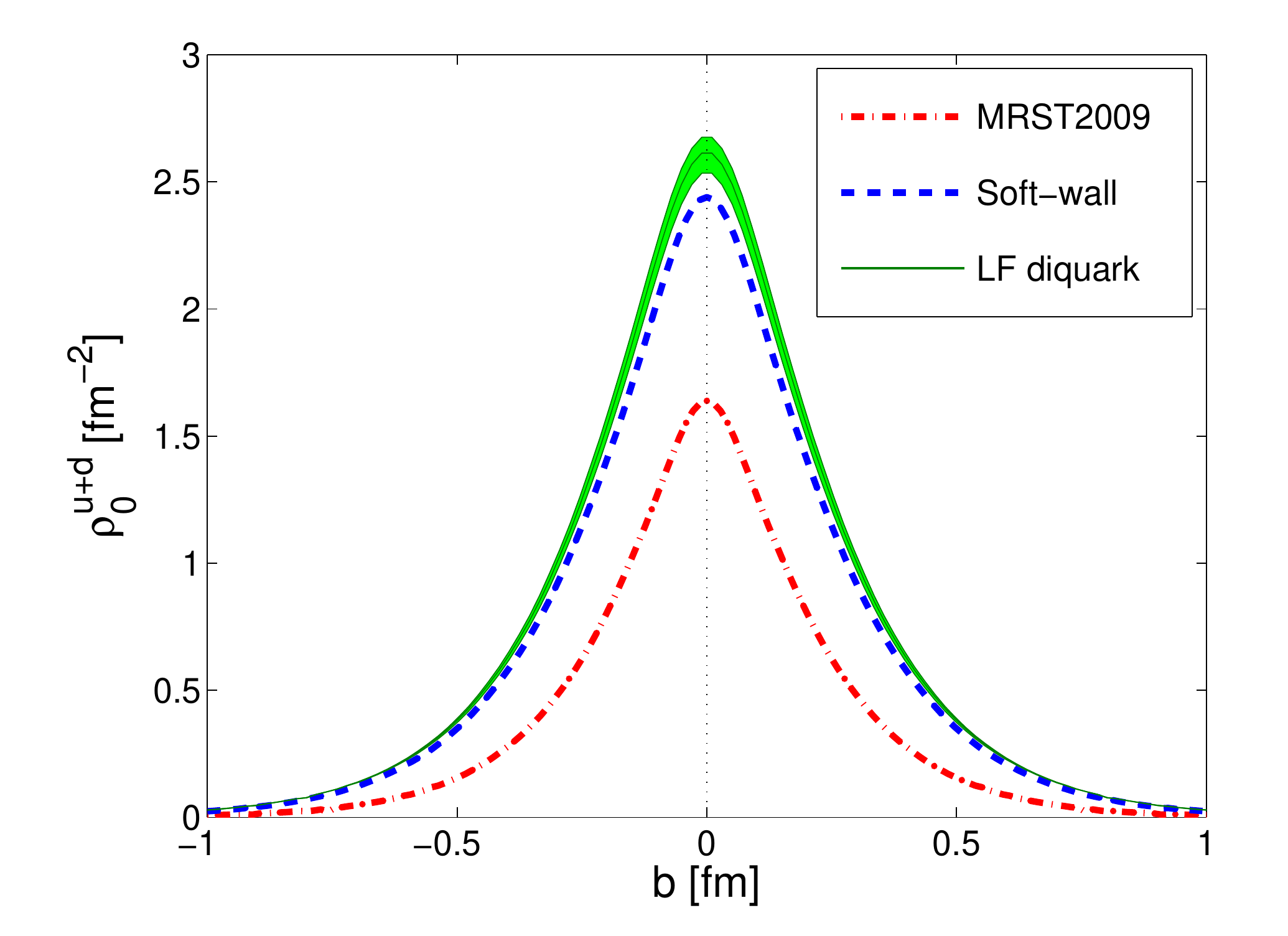}
{(b)}\includegraphics[width=7.5cm,clip]{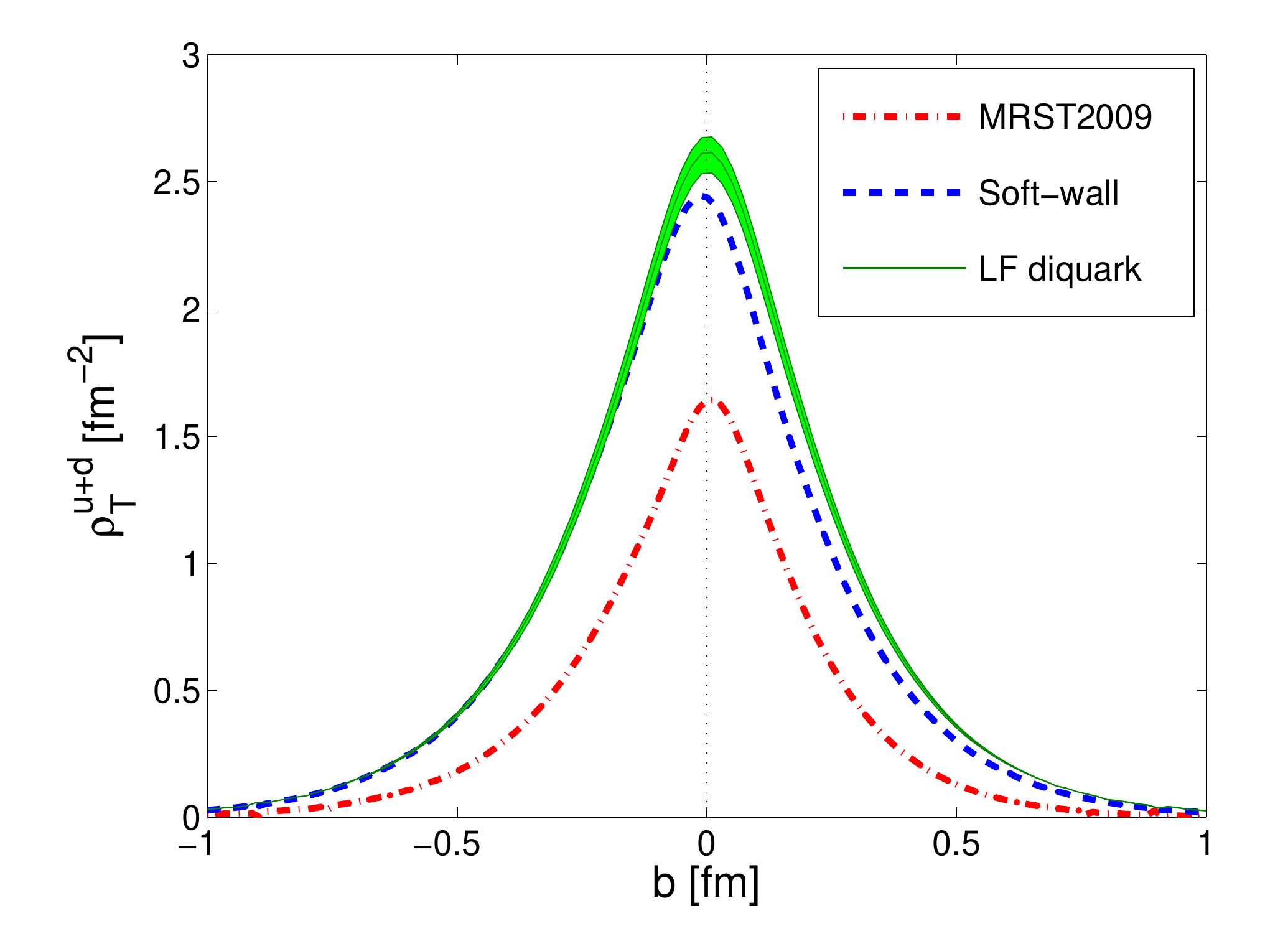}
\end{minipage}
\caption{\label{LMD_combine}Plot of up and down quarks combined longitudinal momentum densities in an unpolarized (left) and in a transversely polarized (right) nucleon.}
\end{figure*}
\begin{figure*}[htbp]
\begin{minipage}[c]{0.98\textwidth}
{(a)}\includegraphics[width=7.5cm,clip]{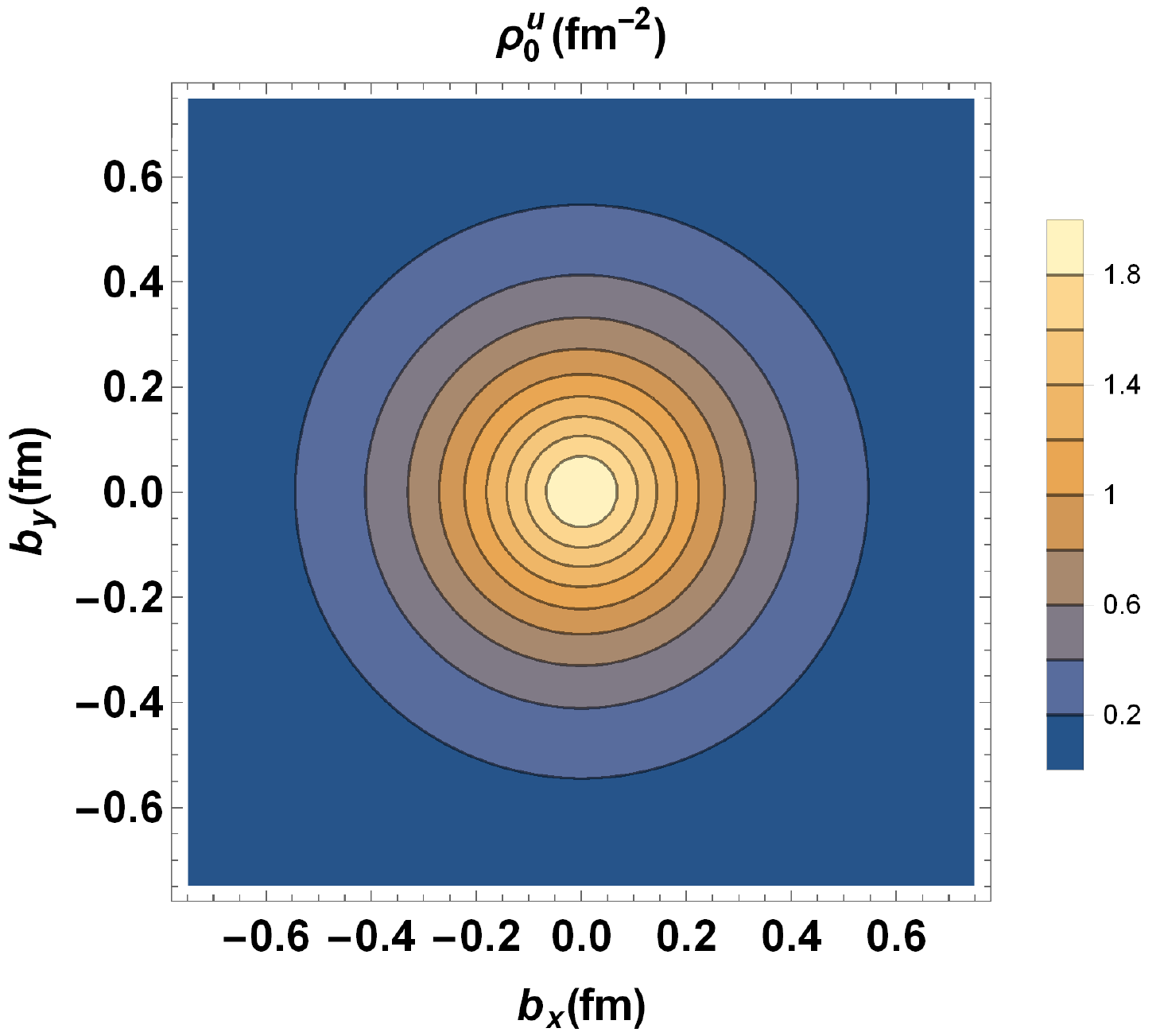}
{(b)}\includegraphics[width=7.5cm,clip]{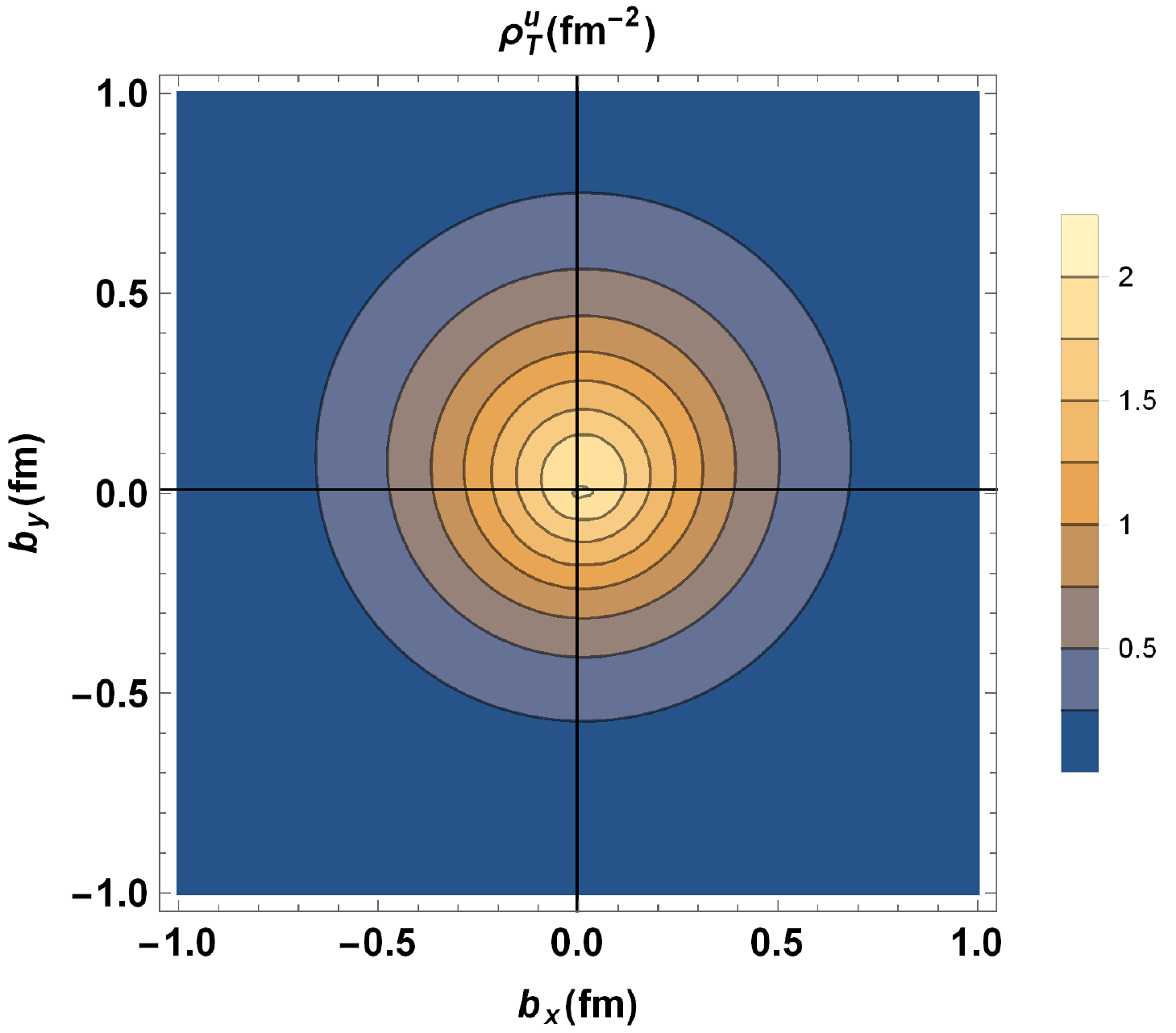}
\end{minipage}
\begin{minipage}[c]{0.98\textwidth}
{(c)}\includegraphics[width=7.5cm,clip]{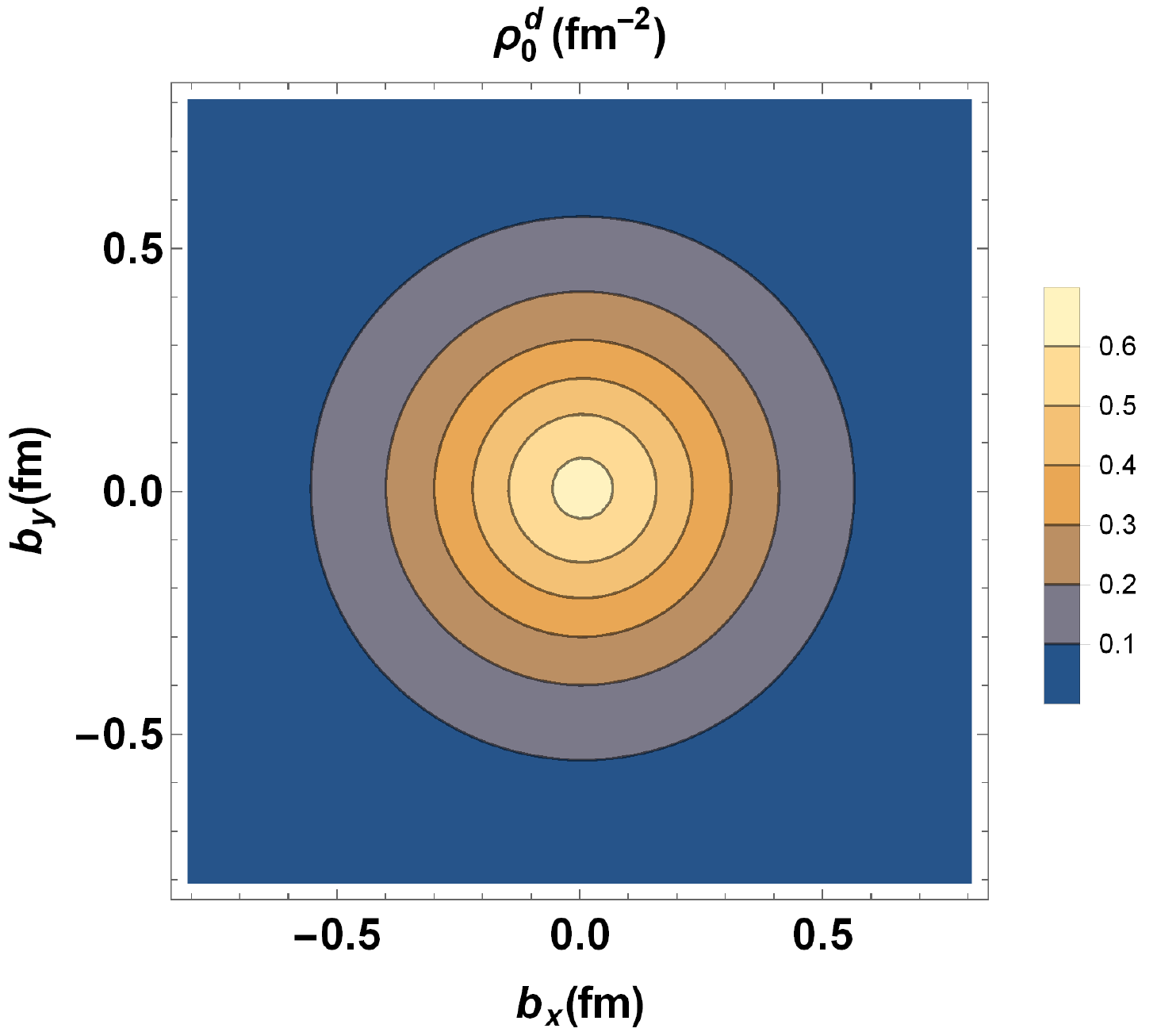}
{(d)}\includegraphics[width=7.5cm,clip]{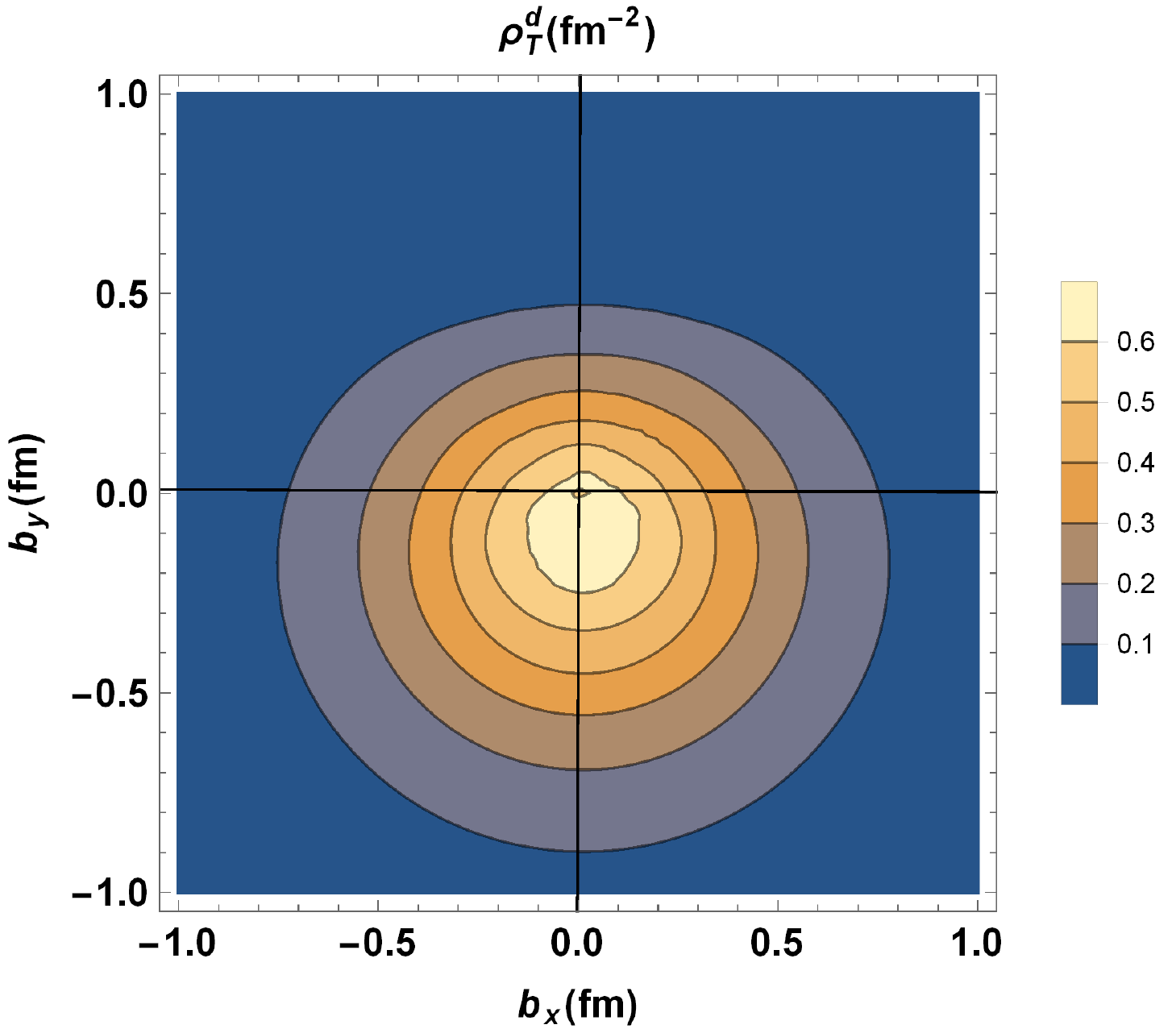}
\end{minipage}
\caption{\label{quarks_contour_LF} Top view of longitudinal momentum densities for up quark (upper panel) and down quark (lower panel) in an unpolarized (left panel) and in a transversely polarized (right panel) nucleon  polarized along $\hat{x}$-direction obtained in LF-diquark model.}
\end{figure*}
\begin{figure*}[htbp]
\begin{minipage}[c]{0.98\textwidth}
{(a)}\includegraphics[width=7.5cm,clip]{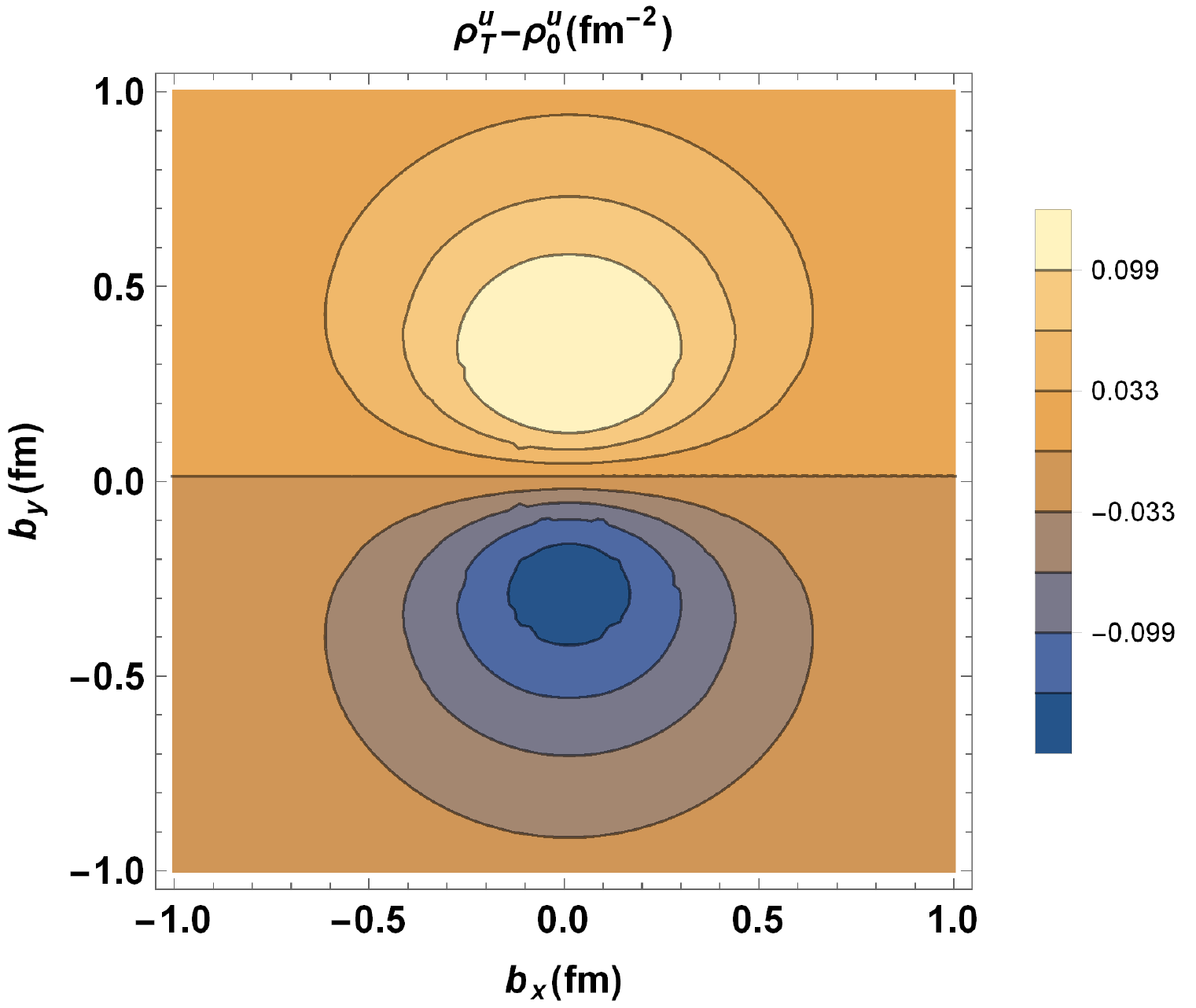}
{(b)}\includegraphics[width=7.5cm,clip]{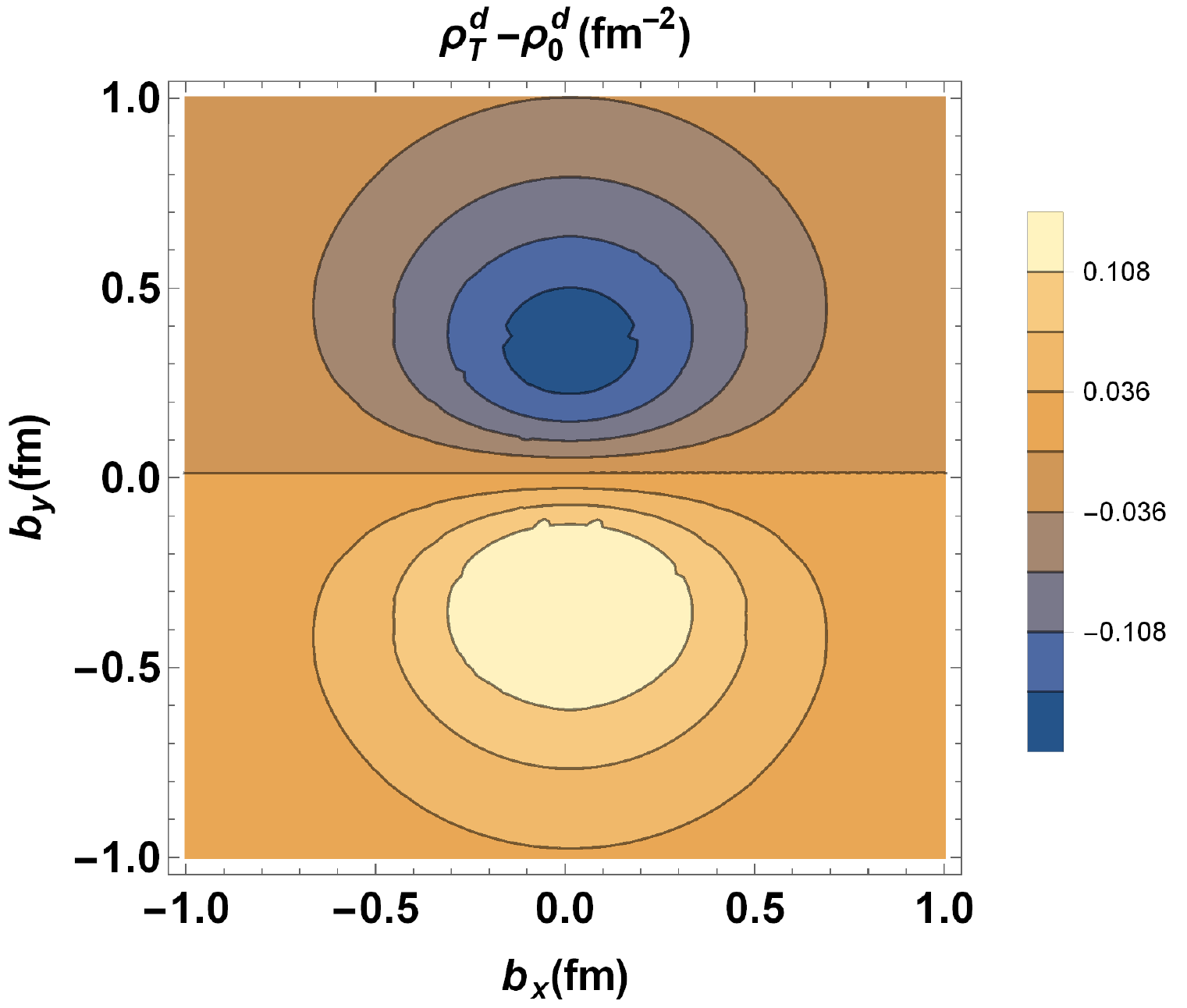}
\end{minipage}
\caption{\label{asymmetry_contour}Plots of longitudinal momentum densities asymmetry $(\rho_T(b)-\rho(b))$ for up quark (left panel) and down quark (right panel) obtained in LF-diquark model.}
\end{figure*}
\begin{figure*}[htbp]
\begin{minipage}[c]{0.98\textwidth}
{(a)}\includegraphics[width=7.5cm,clip]{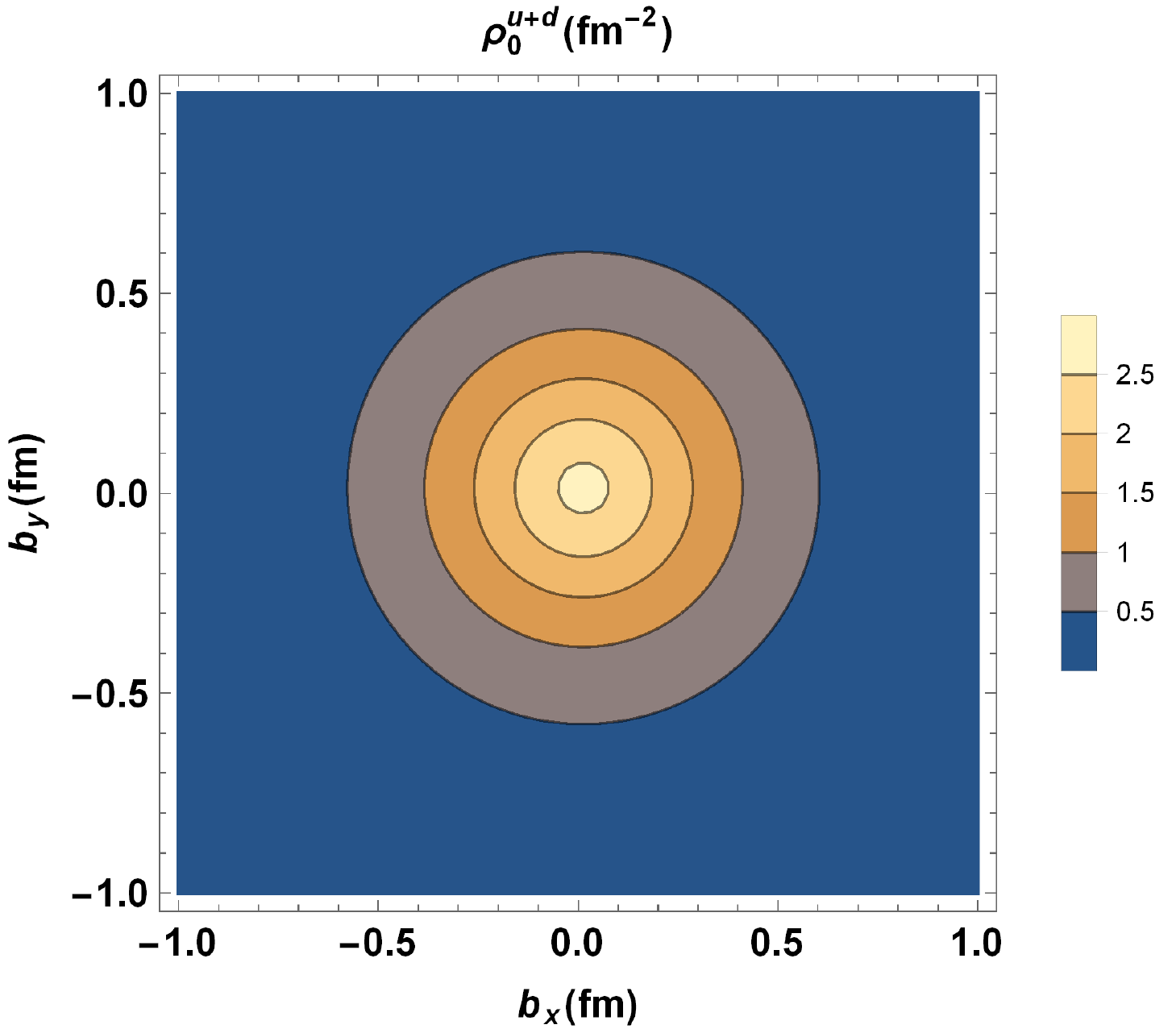}
{(b)}\includegraphics[width=7.5cm,clip]{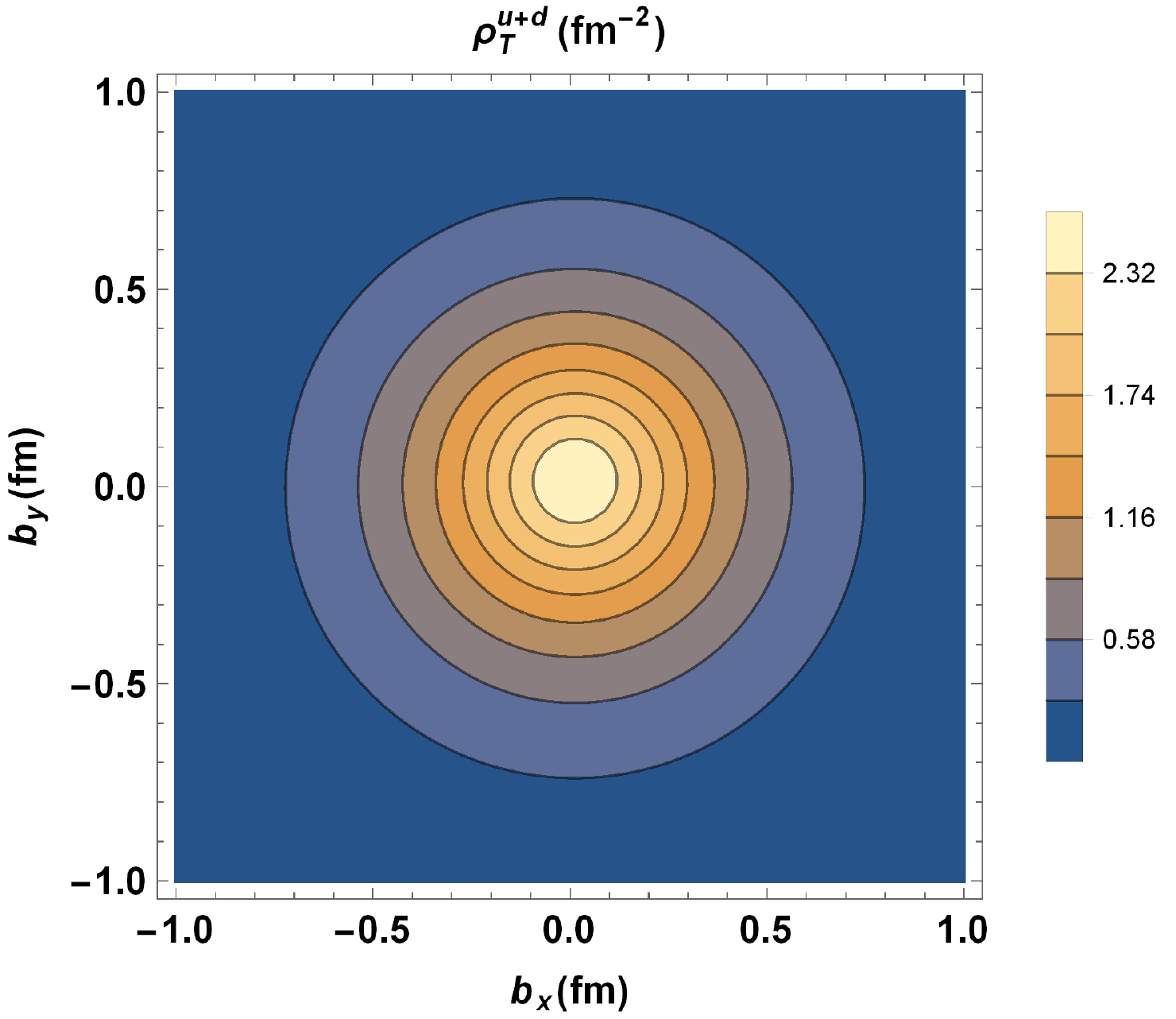}
\end{minipage}
\caption{\label{combine_contour_LF}Plots of longitudinal momentum densities for combination of up and down quark in unpolarized (left panel) and transversely polarized (right panel) nucleon.}
\end{figure*}
\begin{figure*}[htbp]
 \begin{minipage}[c]{0.98\textwidth}
{(a)}\includegraphics[width=7.2cm,clip]{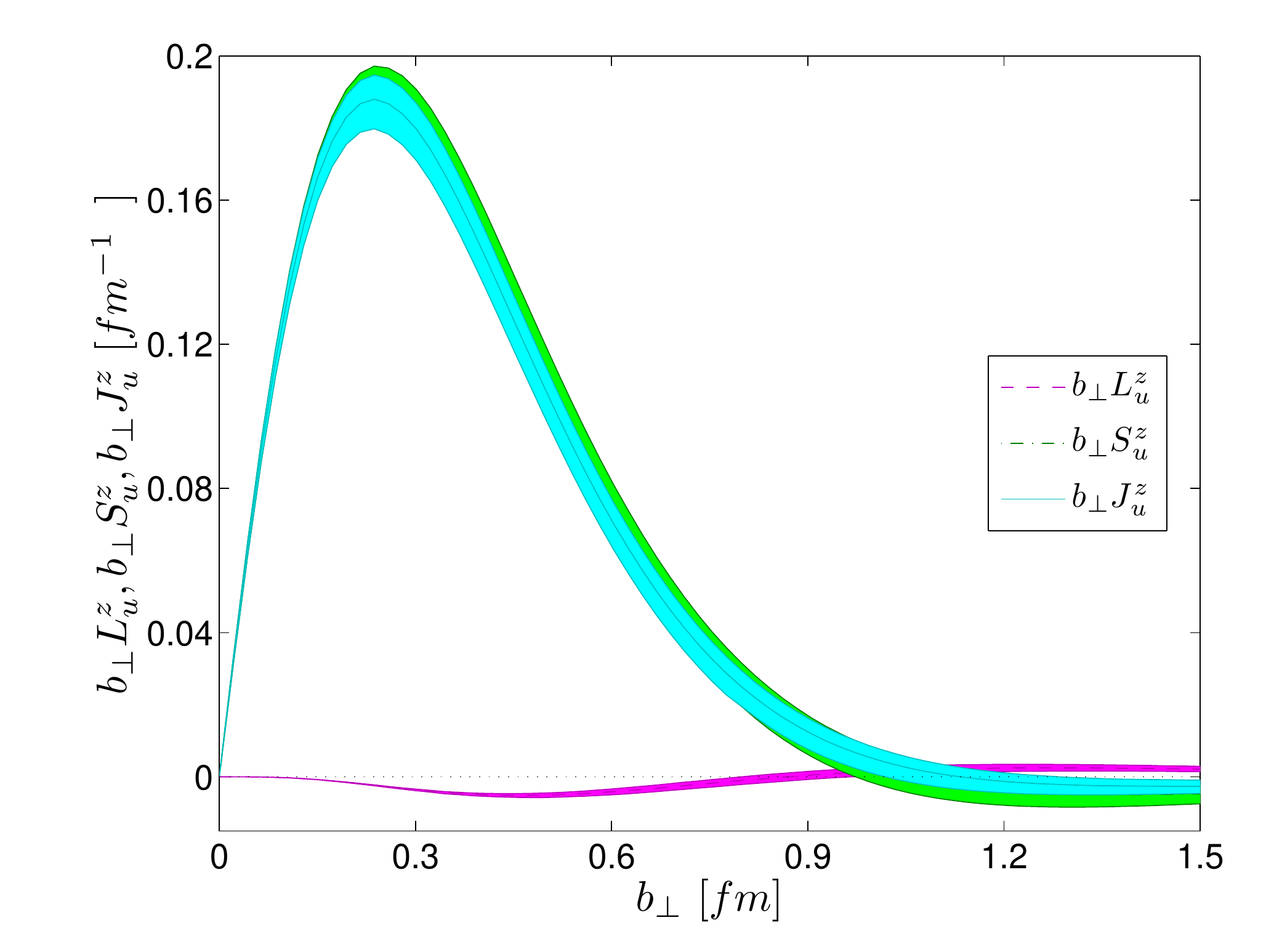}
{(b)}\includegraphics[width=7.5cm,clip]{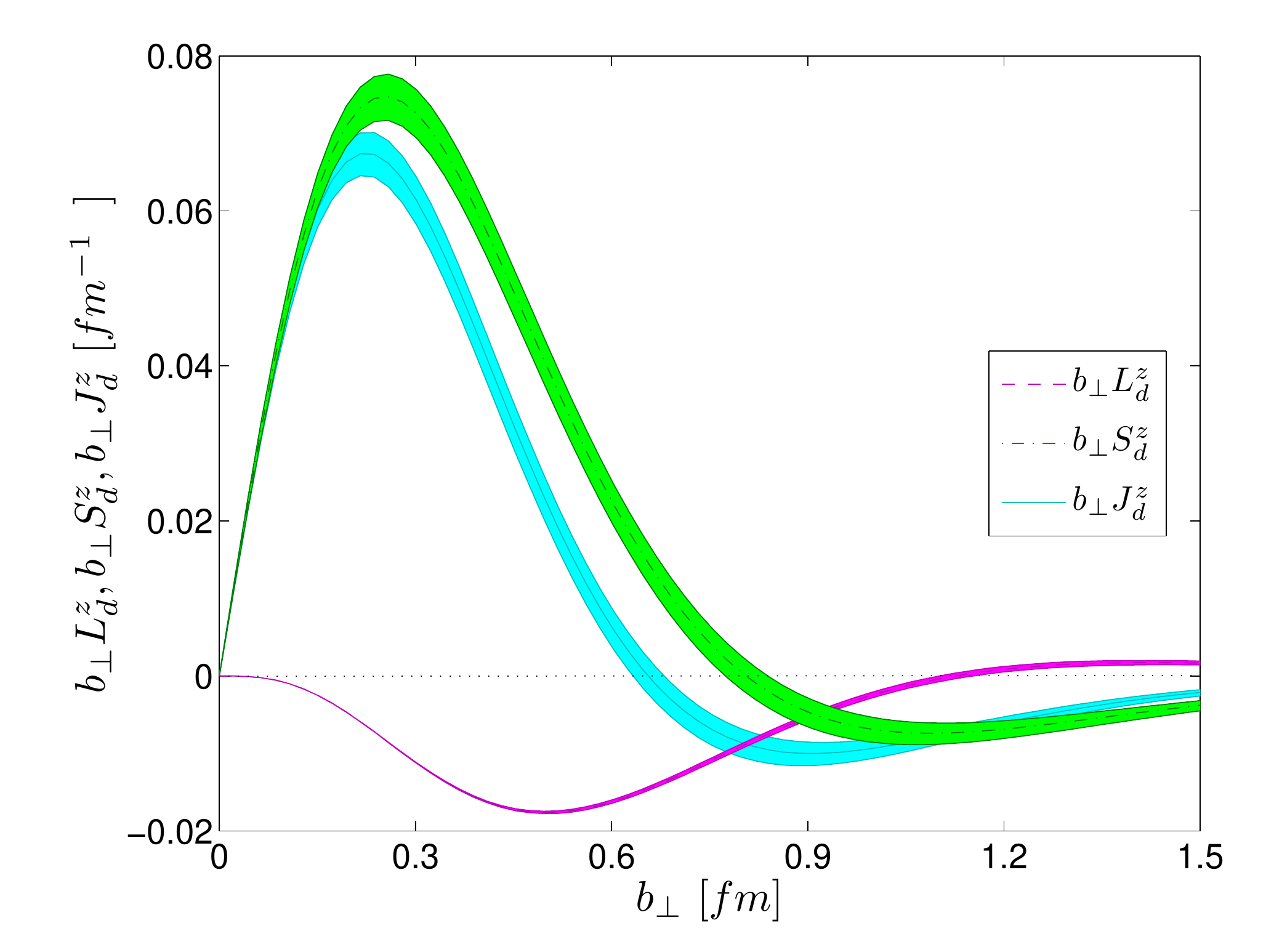}
\end{minipage}
\begin{minipage}[c]{0.98\textwidth}
{(c)}\includegraphics[width=7.2cm,clip]{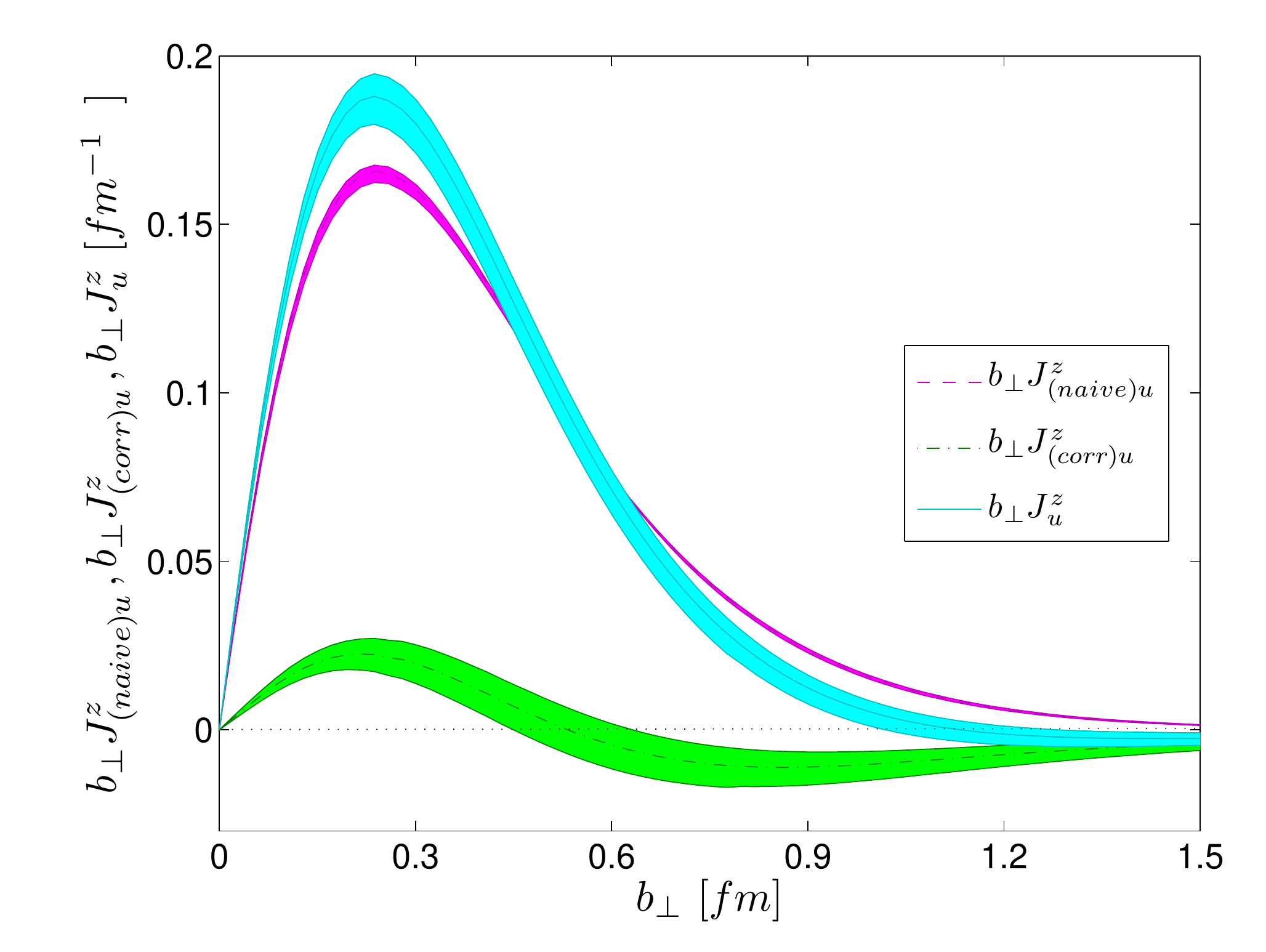}
{(d)}\includegraphics[width=7.5cm,clip]{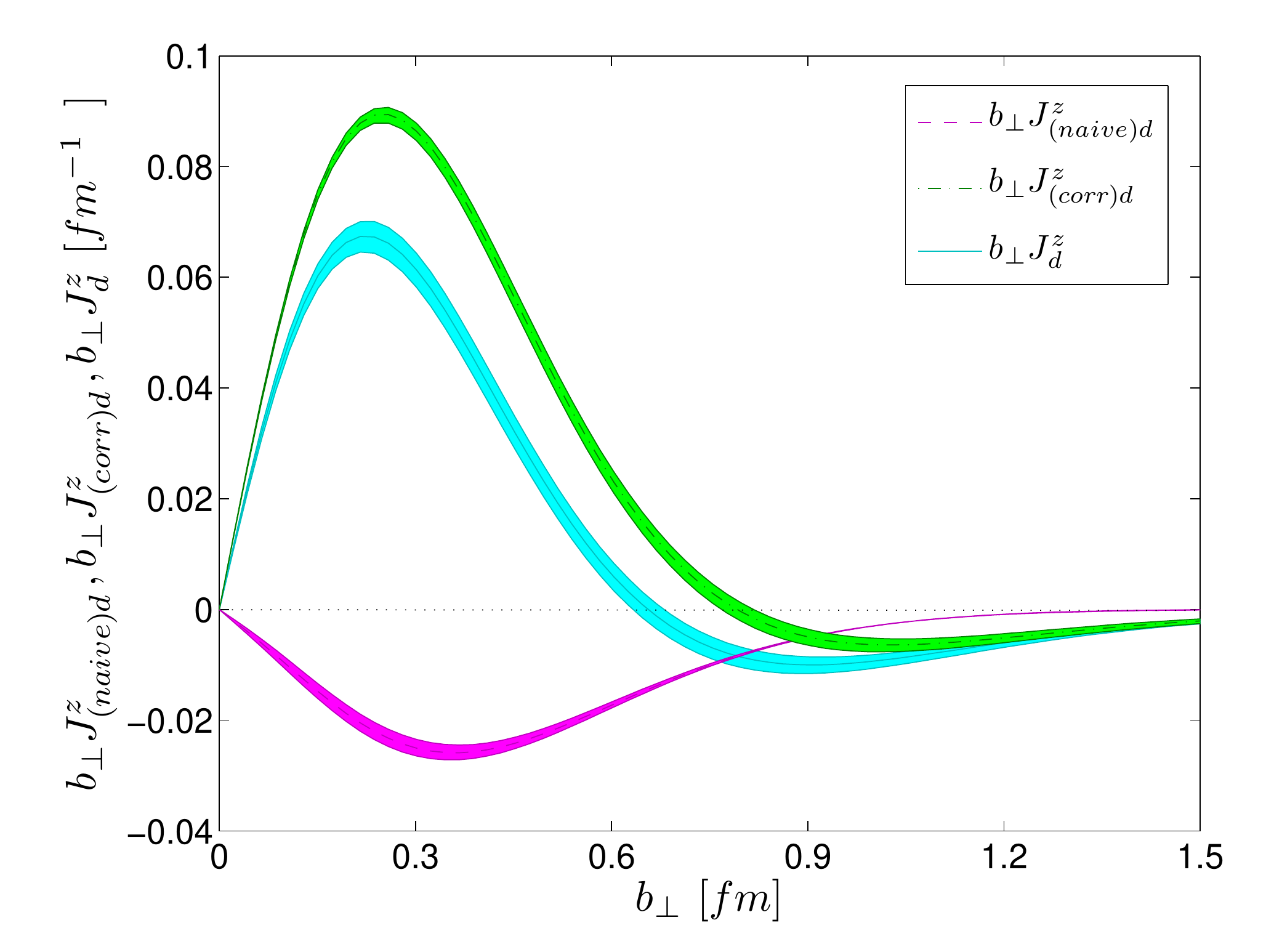}
\end{minipage}
\begin{minipage}[c]{0.98\textwidth}
{(e)}\includegraphics[width=7.2cm,clip]{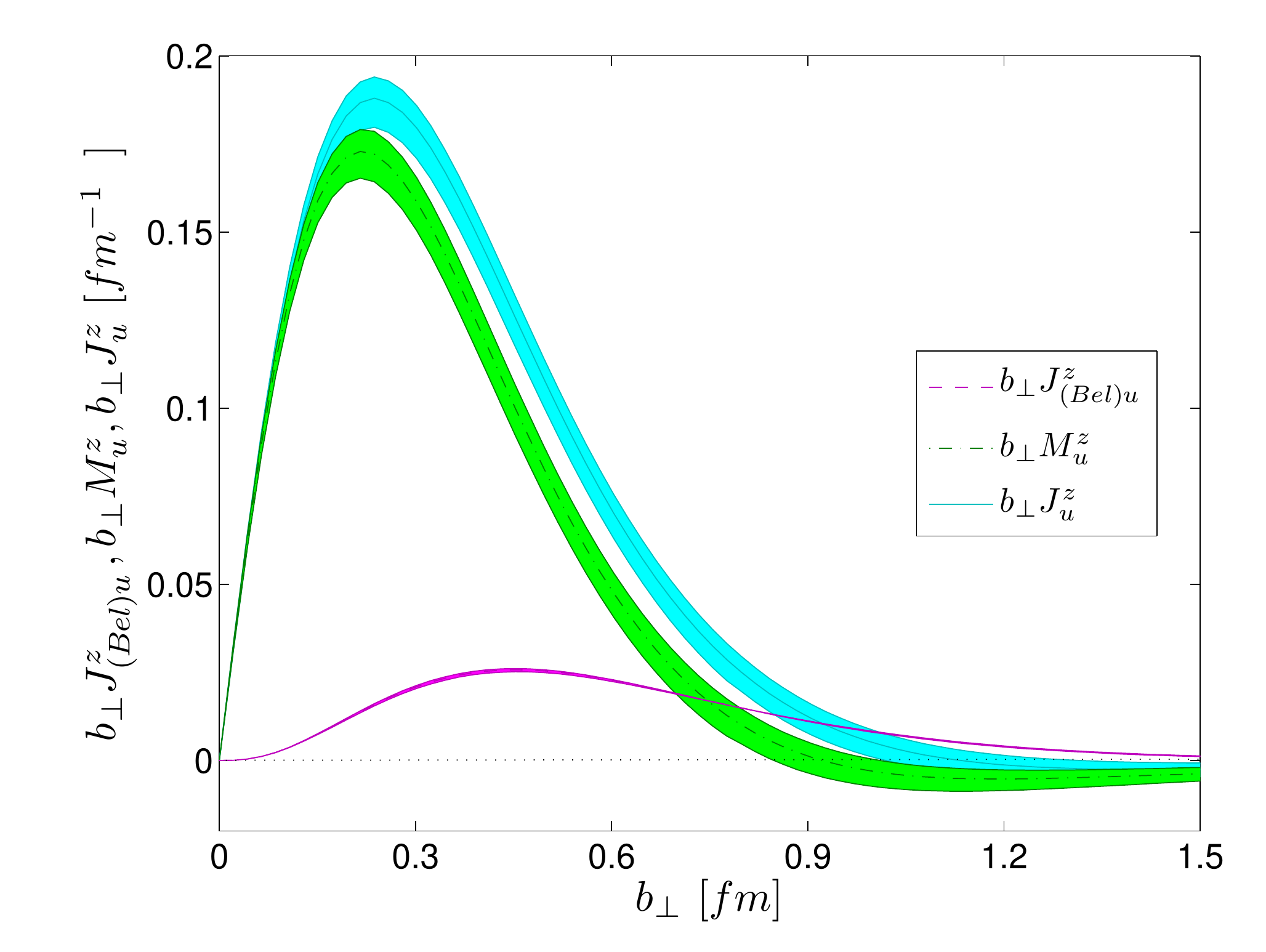}
{(f)}\includegraphics[width=7.5cm,clip]{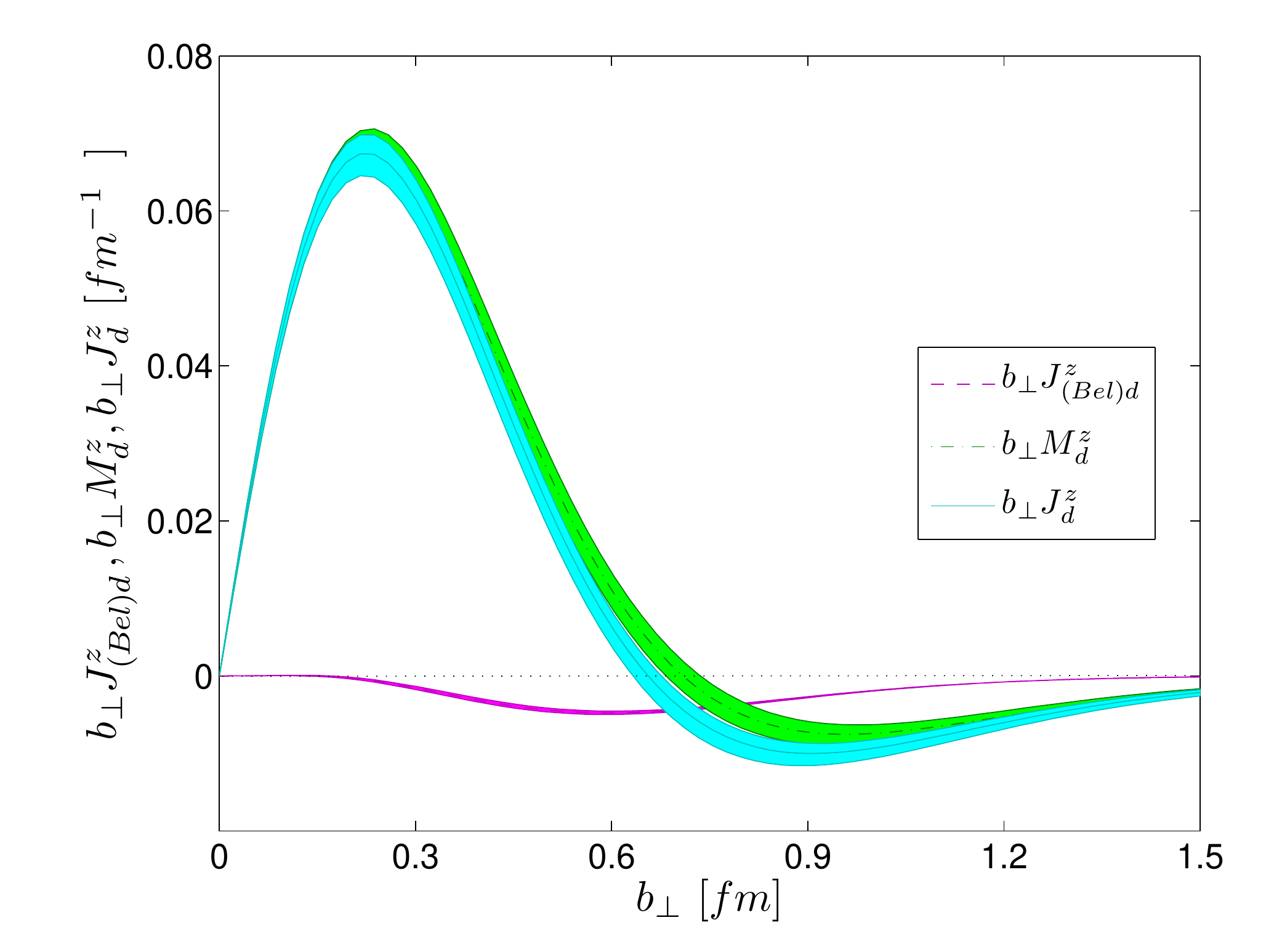}
\end{minipage}
\caption{\label{AMD_plot}Plots of longitudinal angular momentum densities for up quark (left panel) and down quark (right panel) in the transverse plane obtained in LF-diquark model. The upper panel corresponds to the kinetic total angular momentum $J^z(b_{\perp})$ (solid line) as resulting from the sum of kinetic orbital angular momentum $L^z(b_{\perp})$ (dashed line) and spin $S^z(b_{\perp})$ (dot-dashed line). The middle panel corresponds to the kinetic total angular momentum $J^z(b_{\perp})$ (solid line) expressed as the sum of ``naive'' total angular momentum density $J^z_{(naive)}(b_{\perp})$ (dashed line) and the corresponding correction $J^z_{(corr)}(b_{\perp})$ (dashed dot line). Lower panel: the kinetic total angular momentum $J^z(b_{\perp})$ (solid line) resulting from the sum of Belinfante-improved total angular momentum $J^z_{(Bel)}(b_\perp)$ (dashed line) and the total divergence term $M^z(b_\perp)$ (dot-dashed line). The error bands correspond to 2$\sigma$ error in the quark-diquark model parameters.}
\end{figure*}
\begin{figure*}[htbp]
\begin{minipage}[c]{0.98\textwidth}
{(a)}\includegraphics[width=7.2cm,clip]{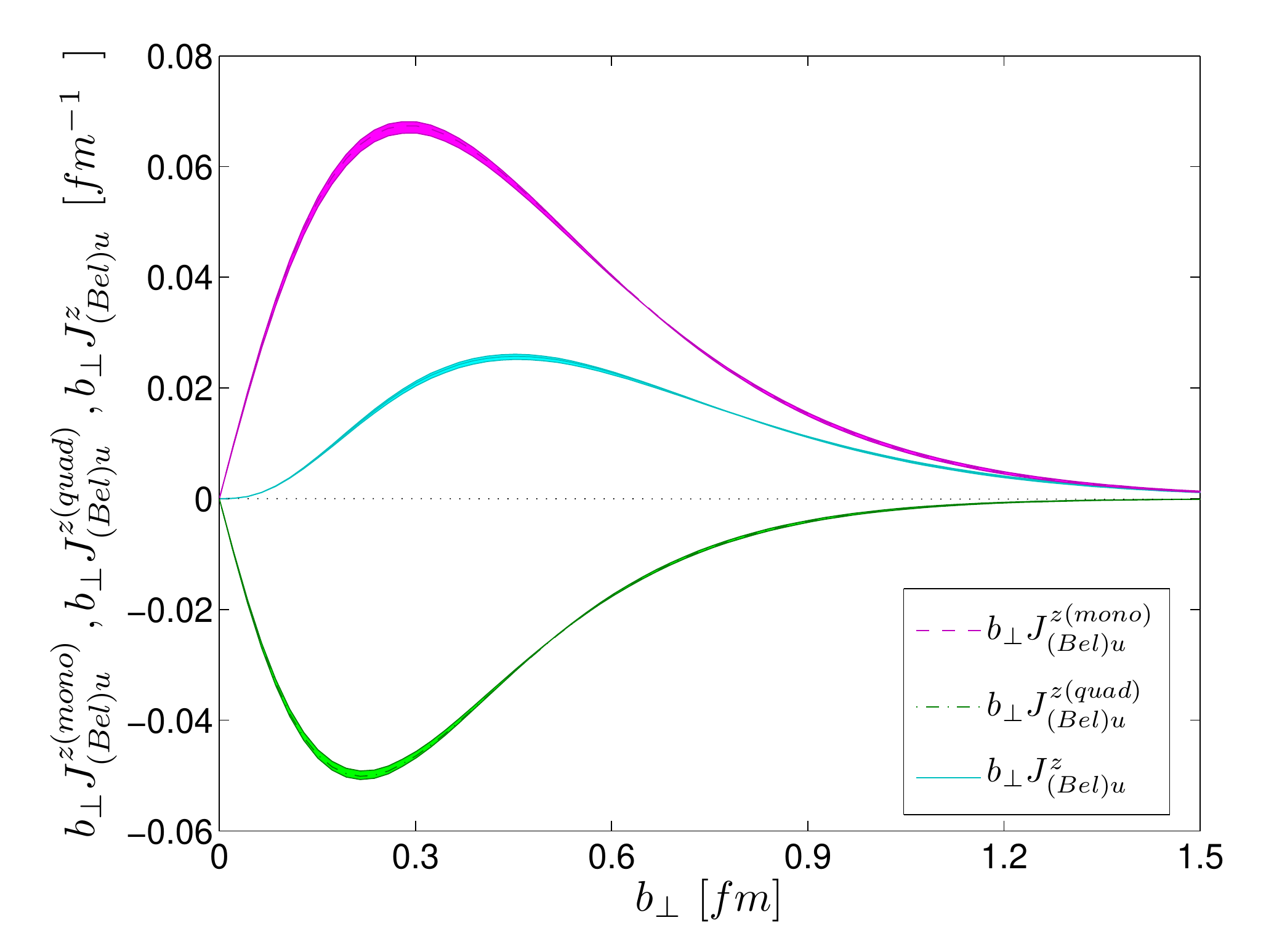}
{(b)}\includegraphics[width=7.5cm,clip]{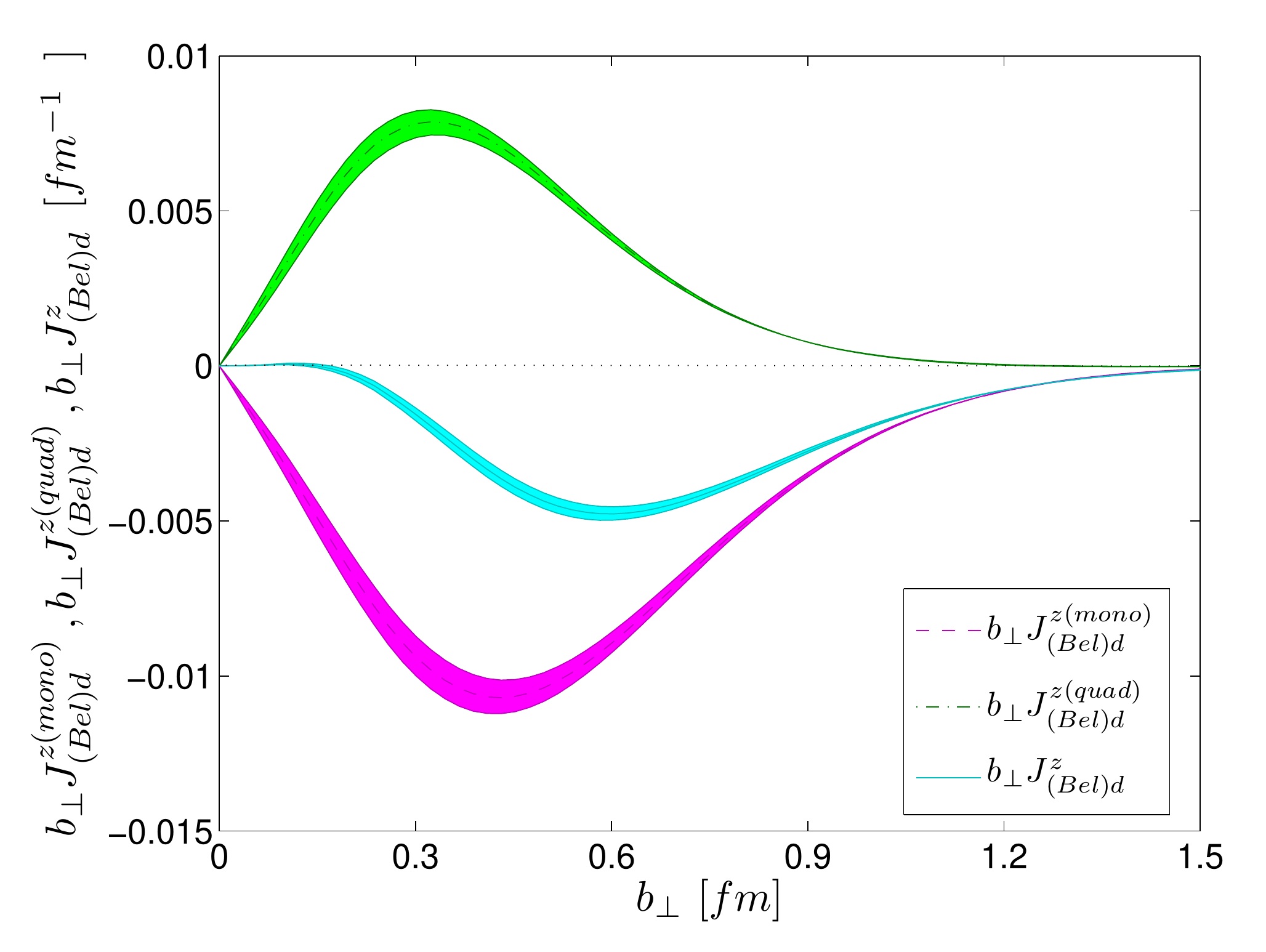}
\end{minipage}
\caption{\label{AMD_plot2}Plots of Belinfante-improved total angular momentum $J^z_{(Bel)}(b_\perp)$ (solid line) expressed as the sum of the monopole term $J^{z(mono)}_{Bel}(b_\perp)$ (dashed line) and the quadrupole term $J^{z(quad)}_{Bel}(b_\perp)$ (dot-dashed line). (a) up quark, (b) down quark. The error bands correspond to 2 $\sigma$ error in the quark-diquark model parameters.}
\end{figure*}
In Fig.\ref{GFFs}(a) and \ref{GFFs}(b), we show the total GFFs $A(Q^2)$ and $B(Q^2)$ for proton depending on different quark and diquark combination. Individual quark and diquark contributions to $A_{tot}(Q^2)$ and $B_{tot}(Q^2)$  are shown in Fig.\ref{GFFs}(c) and \ref{GFFs}(d) respectively. The error bands in the plots correspond to 2$\sigma$ error in the model parameters. It can be noticed that For spin non-flip GFF, although up and down quarks contributions are slightly different,  the contributions from their corresponding diquarks, i.e. $(ud)$ and $(uu)$ are almost the same. Effectively, the $A_{tot}$ for ($u+ud$) and ($d+uu$) are more or less the same. For spin-flip GFF, the contribution of up quark is opposite to down quark and the contributions from the respective diquark are also opposite to each other. This is due to the fact that the anomalous magnetic moment for up quark is positive but it is negative for down quark. One can also notice that at zero momentum transfer, the quark-diquark model satisfies the physical condition i.e. $A_{tot}(0)=1$ and $B_{tot}(0)=0$.

The GFFs can also be obtained from the second Mellin's moment of GPDs. In a recent study \cite{neetika2016}, momentum transfer dependence of GPDs have investigated where the GPDs are obtained from the MRST2009 global fit. GPDs can be extracted from the various phenomenological parametrizations available in the literature. The simplest form of the parametrization for extraction of proton GPDs is a Gaussian form of wavefunction having $x$ and $Q^2$ dependence. At small momentum transfer, a Regge parametrization for GPDs $H(x,Q^2)=q(x)\exp[-\alpha \ Q^2]$ and the modified version $q(x)\exp[-\alpha (1-x) Q^2]$ of the same is used for the analysis at large momentum transfer \cite{neetika2016}. 
The spin non-flip GPD can be written as \cite{neetika2016}
\be
H^q(x,Q^2)= q(x) \exp \Big[- a_q \ \frac{(1-x)^2}{x^{m}} Q^2 \Big],
\ee
where $a_u$, $a_d$ and $m$ are the free fitted parameters from the low $Q^2$ experimental data on the proton form factors. The PDFs measured at NNLO in strong coupling parameter by ``MSTW2009'' are expressed as \cite{MartinA.D.2009}
\bea
x \ u(x) &=& 0.22 \ x^{0.28} (1-x)^{3.36} \nonumber\\&&\times(1+ 4.43\ \sqrt{x}+38.6 \ x),\nonumber\\
x \ d(x) &=& 17.94 \ x^{1.08} (1-x)^{6.15}\nonumber\\&&\times (1- 3.64 \ \sqrt{x}+ 5.26 \ x).
\eea
The spin-flip GPD $E^q(x, Q^2)$ is given by
\be
E^q(x,Q^2)=  \mathcal{E}_q(x) \exp \Big[- a_q \ \frac{(1-x)^2}{x} Q^2 \Big], 
\ee
with
\bea
\mathcal{E}_u(x) &=& \frac{\kappa_u}{N_u} (1-x)^{\kappa_1} u(x),\nonumber\\
\mathcal{E}_d(x) &=& \frac{\kappa_d}{N_d} (1-x)^{\kappa_2} d(x),
\eea
where the normalization of up and down quark GPDs to their corresponding anomalous magnetic moments $\kappa_u=1.673, \kappa_d=-2.033$ leads to the $k_1=1.53, k_2=0.31, N_u=1.52 \ \text{and} \ N_d=0.95$. We adopt all the input parameters from Ref. \cite{neetika2016}. 
In Fig.\ref{GFFs_AB}, we compare the GFFs $A(Q^2)$ and $B(Q^2)$ for up and down quarks evaluated in the quark-diquark model with the results obtained in a soft-wall AdS/QCD model \cite{Sharma2014} and the parametrization of GPDs based on latest global analysis by ``MRST2009"\cite{neetika2016}. We find that for both quarks, the quark-diquark model is in good agreement with the soft-wall model whereas for up quark, the diquark model deviates a little from the parametrization of GPDs. 
\section{Longitudinal momentum densities}\label{lmd}
Transverse charge and magnetization densities \cite{Miller2007,Carlson2008} are defined as the two dimensional Fourier transformation of the Dirac $F_1(q^2)$ and Pauli $F_2(q^2)$ form factor with respect to momentum transferred $q_\perp$. Charge and magnetization densities in transverse plane for nucleon have been studied in various phenomenological models \cite{Mondal2016,Chakrabarti2014,Sharma2014,MondalChandan2015,Kumar2014}. 
Similar to charge densities, one can also evaluate the $p^+$ densities in the transverse plane by taking the two dimensional Fourier transform of the GFF $A(q^2)$. The $T^{++}$ component of the energy-momentum tensor gives to the longitudinal momentum $P^+$  as
\begin{equation}
P^+= \int T^{++} \ d^2x_\perp \ dx_+,
\end{equation}
and the GFFs are connected to the matrix element of $T^{++}$, thus 
one can interpret the two-dimensional FT of the GFF $A(Q^2)$ as the longitudinal momentum density in the transverse plane \cite{Abidin2008}.
\bea
\rho(b_\perp)&=& \int \frac{d^2 \mb{q}_\perp}{(2 \pi)^2} A(q^2) e^{-i \mb{q}_\perp \mb{b}_\perp}\nonumber\\&=& \int \frac{q_\perp \ dq_\perp}{2 \pi} J_0(|q_\perp| |b_\perp|) A(q^2),
\eea
where $b= |b_\perp|$ is the impact parameter and $J_0$ is the Bessel function of zeroth order. The addition of spin-flip matter form factor $B(q^2)$ to unpolarized quark density modify when one considers a transversely polarized nucleon. Therefore for a polarized nucleon, modified quark density can be written as
\bea
\rho_T(b_\perp)&=& \rho(b)+ \sin(\phi_b -\phi_s) \nonumber\\
&&\times\int_{0}^{\infty} \frac{dq_\perp \ q_\perp^2}{4 \pi} \frac{J_1(|q_\perp| \ |b_\perp|)}{M_n} B(q^2),
\label{asymmetry}
\eea
where $M_n$ is the mass of nucleon. The transverse polarization of the nucleon is denoted by $S_\perp= \cos\phi_s \ \hat{x} +\sin\phi_s \ \hat{y}$ and transverse impact parameter is denoted by $b_\perp= b ( \cos\phi_b \ \hat{x} + \sin\phi_b \ \hat{y})$. The second part in Eq.(\ref{asymmetry}), adds the deviation from the circular symmetry of the unpolarized density. In Fig.\ref{LMD}(a) and Fig.\ref{LMD}(b), we present the results for the longitudinal momentum densities for up quark in an unpolarized and in a transversely polarized proton respectively. We compare the results of LF-diquark model with the ``MRST2009'' parametrization and the soft-wall AdS/QCD model. We observe that for the unpolarized nucleon, results are axially symmetric, however the magnitude of LF-diquark model is larger than the ``MRST2009'' and soft-wall AdS/QCD model.
In the case of transversely polarized nucleon, the distribution gets distorted due to addition of $B(q^2)$ in all cases. In Fig. \ref{LMD}(c) and \ref{LMD}(d), we show the longitudinal momentum densities for down quark. It can be noticed that for the unpolarized case, distribution is axially symmetric and widely spread as compare to distribution for up quark whereas in transversely polarized the distribution is distorted. In Fig. \ref{LMD_combine}, we present the sum of up and down quark distributions in an unpolarized and a transversely polarized proton. One can find that for all cases the unpolarized distributions almost overlap with the distributions for transversely polarized proton. This is due to the fact that the distortions for up and down quark in a transversely polarized nucleon have opposite directions. Further in Fig. \ref{quarks_contour_LF}(a) and \ref{quarks_contour_LF}(c), we show the top view of three dimensional distributions in the impact-parameter space for up and down quarks in an unpolarized nucleon. From these plots it is clear that the distribution is axially symmetric. The magnitude of up quark is larger than that of down quark. A similar plots for up and down quark in transversely polarized nucleon are shown in Fig. \ref{quarks_contour_LF}(b) and \ref{quarks_contour_LF}(d). The distributions get distorted due to the additional contribution coming from $B(q^2)$ to symmetric contribution. The distortion in down quark distribution is stronger compared to up quark and effectively exhibits a dipolar  structure. By removing the axially symmetric part from density $\rho_T(b)$ i.e., $(\rho_T(b)-\rho(b))$, we obtain the angular dependent part of the density which gives the dipole pattern and is clearly reflected in Fig. \ref{asymmetry_contour}(a) and \ref{asymmetry_contour}(b) for up and down quark. The top view of the longitudinal momentum density for combined up and down quark in the impact-parameter space for unpolarized and transversely polarized nucleon is shown in Fig. \ref{combine_contour_LF}(a) and \ref{combine_contour_LF}(b) respectively. It shows that in a unpolarized nucleon, distribution is axially symmetric but a little distortion is appeared for transversely polarized nucleon.In comparison with scalar diquark model \cite{Chakrabarti2015}, it has been observed that the qualitative nature of gravitational form factor $A(Q^2)$ in both the scalar and vector diquark models is the same, thus, for an unpolarized nucleon the longitudinal distributions for both $u$ and $d$ quark also show the same qualitative nature. But the magnitude in the vector diquark model is quite large as compared to scalar diquark model. However, the result in scalar diquark model is in more or less in agreement with the phenomenological parametrization of GPDs whereas the result in vector diquark model agrees well with AdS/QCD model \cite{Sharma2014}.
For a transversely polarized nucleon, the longitudinal distribution of both $u$ and $d$ quark in scalar diquark model gets shifted along the same direction (positive $\hat{y}$-direction for nucleon polarized along positive $\hat{x}$-direction). However, for the vector diquark model, the shifting of the distributions for $u$ quark is opposite to the $d$ quark. Since, the gravitational form factor $B(Q^2)$ in  vector diquark model for u quark is positive but it is negative for d quark which is responsible  for the opposite shifting of $u$ and $d$ quark distribution. But, in the case of scalar diquark model the gravitational form factor $B(Q^2)$ for both $u$ and $d$ quark are roughly the same, which affects the shifting of distribution in the same direction.
\section{Angular momentum distributions}\label{amd}
Ji has shown that the total angular momentum of quarks and gluons can be expressed by the sum rule $J^{q/g}(0)=\frac{1}{2}(A^{q/g}(0)+B^{q/g}(0))$. For $q^2\neq 0$, one cannot get the correct form of quark angular momentum distribution in impact-parameter space by the two dimensional Fourier transformation of only $J^q(q^2)$ \cite{Leader2014,LiuKeh-Fei2016,Polyakov2003}. In Ref.\cite{Adhikari2016}, different definitions of the angular momentum density have been compared and it has been concluded that none of the definitions agree at the density level. Recently, Lorc$\acute{e}$ $et.~ al.$ \cite{Lorce2017} have idenfied all the missing terms and explicitly showed that no discrepancies are found between the different definitions of angular momentum. The spatial distributions of orbital angular momentum and spin inside the nucleon are defined as \cite{Lorce2017},
\bea
 L^{z}(b_\perp)
&=& s^z\int\frac{\ud^2\boldsymbol{q}_\perp}{(2\pi)^2}\,e^{-i\mb{q}_\perp\cdot\mb{b}_\perp}\left[L(t)+t\,\frac{\ud L(t)}{\ud t}\right]_{t=-\mb q^2_\perp},\label{L2D}\\
S^{z}(b_\perp)&=&\frac{s^z}{2}\int\frac{\ud^{2}\boldsymbol{q}_\perp}{(2\pi)^2}\,e^{-i\mb{q}_\perp\cdot\mb{b}_\perp}\,G_{A}(-\mb q^2_\perp) ,
\eea
where $L(t)$ is given by 
\bea
L(t)=\frac{1}{2}\left[A(t)+B(t)-G_A(t)\right],
\eea
and $G_A(t)$ is the axial vector form factor (see in appendix A). Defining the 2D Fourier transform of the form factors as
\begin{equation}
\tilde{F}(b_\perp)=\int\frac{\ud^2 \boldsymbol{q}_\perp}{(2\pi)^2}\,e^{-i\mb q_\perp\cdot\mb b_\perp}\, F(-\mb q^2_\perp) \; , \label{fou2d}
\end{equation}
one can write 
\bea
 L^{z}(b_\perp)&=&-\frac{s^z}{2}\,b_{\perp}\,\frac{\ud \tilde{L}(b_{\perp})}{\ud b_{\perp}},\quad
S^{z}(b_\perp)=\frac{s^{z}}{2}\,\tilde{G}_{A}(b_{\perp}) \; ,
\eea
Similarly, one defines the impact-parameter densities of Belinfante-improved total angular momentum and total divergence \cite{Lorce2017},
\bea
J^{z}_\text{Bel}({b}_\perp)&=s&^z\int\frac{\ud^2\boldsymbol{q}_\perp}{(2\pi)^2}\,e^{-i\mb{q}_\perp\cdot\mb{b}_\perp}\left[J(t)+t\,\frac{\ud J(t)}{\ud t}\right]_{t=-\mb q^2_\perp}\nonumber\\
&=&-\frac{s^z}{2}\,b_{\perp}\,\frac{\ud \tilde{J}(b_{\perp})}{\ud b_{\perp}} \; ,\\
 M^{z}({b}_\perp)
&=&-\frac{s^z}{2}\int\frac{\ud^{2}\boldsymbol{q}_\perp}{(2\pi)^2}\,e^{-i\mb{q}_\perp\cdot\mb{b}_\perp}\left[t\,\frac{\ud G_{A}(t)}{\ud t}\right]_{t=-\mb q^2_\perp}\nonumber\\
&=&\frac{s^z}{2}\left[\tilde{G}_{A}(b_{\perp})+\frac{1}{2}\,b_{\perp}\frac{\ud\tilde{G}_A(b_{\perp})}{\ud b_{\perp}}\right] \; ,\label{divb}
\eea
where
\bea
J(t)=\frac{1}{2}\left[A(t)+B(t)\right].
\eea
The Belinfante-improved total angular momentum can also be expressed as the sum of monopole and quadrupole contributions:
\bea
 J_{\text{Bel}}^{z(\text{mono})}(b_{\perp})&=\frac{s^{z}}{3}\left[\tilde{J}(b_{\perp})-b_{\perp}\,\frac{\ud\tilde{J}(b_{\perp})}{\ud b_{\perp}}\right] \;,\label{lbur} \\ 
 J_{\text{Bel}}^{z(\text{quad})}(b_{\perp})&=-\frac{s^{z}}{3}\left[\tilde{J}(b_{\perp})+\frac{1}{2}\,b_{\perp}\,\frac{\ud \tilde{J}(b_{\perp})}{\ud b_{\perp}}\right] \; .\label{lquadrup}
\eea
The impact-parameter density of kinetic total angular momentum $\langle J^{z}\rangle({b}_\perp)$ is thus given by
\bea
J^z(b_\perp)=L^{z}({b}_\perp)+ S^{z}({b}_\perp)=J^{z}_\text{Bel}({b}_\perp)+ M^{z}({b}_\perp). 
\eea
$J^z(b_\perp)$ is different form the ``naive'' density: $J^z_{\text{naive}}(b_{\perp})=s^z\tilde{J}(b_{\perp})$ by a correction term
\bea
J^{z}_{\text{corr}}({b}_\perp)=-s^z\left[\tilde{L}(b_{\perp})+\frac{1}{2}\,b_{\perp}\,\frac{\ud \tilde{L}(b_{\perp})}{\ud b_{\perp}}\right] \; .
\label{corrb}
\eea

In Fig.\ref{AMD_plot}, we show the angular momentum densities for up and down quarks for a longitudinally polarized target $\mb{s}=(0,0,1)$ obtained in the light-front quark-diquark model inspired by AdS/QCD. Here we consider all the different definitions described above.  In Fig.\ref{AMD_plot}(a) and (b), we present the kinetic total angular momentum $\langle J^z\rangle(b_\perp)=\langle L^z\rangle(b_\perp)+\langle S^z\rangle(b_\perp)$ as the sum of kinetic orbital and spin contributions for up and down quarks respectively. In this model, the contribution in $J^z(b_\perp)$ from $S^z(b_\perp)$ is stronger than that from $L^{z}(b_\perp)$ for both up and down quarks. $S^z(b_\perp)$ is positive for up quark whereas at higher $b_\perp$ it is negative for $d$ quark. $L^z(b_\perp)$ is negative for both quarks. $J^z(b_\perp)$ effectively shows a positive distribution for up quark but it is positive at low $b_\perp$ and negative at higher $b_\perp$ for down quark. In Fig.\ref{AMD_plot}(c) and (d), the kinetic total angular momentum $ J^z(b_\perp)$ has been compared with the naive density $\tilde{J}(b_{\perp})$ for up and down quarks. The naive density $\tilde{J}(b_{\perp})$ is positive for up quark and negative for down quark. The difference between them is given by the correction term $ J^z_{\text{corr}}(b_\perp)$ in Eq.~\eqref{corrb}. The correction term for up quark is small but for down quark it is large and opposite of $\tilde{J}^z_d(b_{\perp})$. In Fig.\ref{AMD_plot}(e) and (f), we compare the kinetic total angular momentum  $ J^z(b_\perp)$ with the Belinfante-improved total angular momentum $ J^z_\text{Bel}(b_\perp)$. The difference is being attributed to the $ M^z(b_\perp)$ term in Eq.~\eqref{divb}. $ J^z_\text{Bel}(b_\perp)$ is very small compared to $ J^z(b_\perp)$ for both up and down quark and  $ J^z_\text{Bel}(b_\perp)$ is positive for up quark but it is opposite for down quark. In both quark the major contribution in $ J^z(b_\perp)$ is coming from $ M^z(b_\perp)$. The decomposition of the Belinfante-improved total angular momentum $J^z_\text{Bel}(b_\perp)=J_\text{Bel}^{z(\text{mono})}(b_\perp)+J_\text{Bel}^{z(\text{quad})}(b_\perp)$ into its monopole and quadrupole contributions in shown in Fig.\ref{AMD_plot2}. The monopole and quadrupole contributions are opposite to each other and the monopole contribution is larger than the quadrupole contribution for both quarks. It has been shown in Ref.\cite{Lorce2017} that  the integrations of the total divergence term $\langle M^z\rangle(b_\perp)$, the correction term $\langle J^z\rangle_\text{corr}(b_\perp)$ and the quadrupole contribution $\langle J^z_\text{Bel}\rangle_\text{quad}(b_\perp)$ are zero. But at the density level of angular momentum, they need to be taken into account in comparison of different definitions. 
\section{Conclusion}\label{conl}
 We presented the GFFs and longitudinal momentum densities for proton in a light-front quark-diquark model of nucleon. The model is consisting of both scalar and axial-vector diquark and light-front wavefunctions are modeled from the soft-wall AdS/QCD correspondence. The individual contributions from quark and diquark to the proton GFFs have been demonstrated depending on different flavors. The GFFs and longitudinal momentum distributions for up and down quarks are compared with the consequences of a soft-wall AdS/QCD model and a parametrization of GPDs based on the functional form of quark distributions of the latest global analysis by  ``MRST 2009". We found that the light-front diquark model is in better agreement with the soft-wall AdS/QCD model compared to the GPDs parametrization.  We also presented the total angular momentum distributions for up and down quark in impact-parameter space. In this model,
 we illustrated explicitly that no discrepancies are found between the different definitions of angular momentum. In particular, we checked explicitly that the canonical and kinetic angular momentum do coincide at the density level.
\\
\\
\\
\\
{\bf Acknowledgements:}
NK would like to thank Science and Engineering Research Board a statutory board under Department of Science and Technology, Government of India (Grant No. PDF/2016/000722) for financial support under the National Post-Doctoral Fellowship. This work of CM is supported by the funding from the China Postdoctoral Science Foundation under the Grand No. 2017M623279.
NS acknowledges the financial support provided by the University Grants Commission (UGC), Government of India (Grant No. F.4-2/2006 (BSR)/PH/16-17/0007), under the Dr. D. S. Kothari postdoctoral fellowship scheme.

\appendix
\section{Axial vector form factor}
The axial vector form factor $G_A(Q^2)$ in terms of the overlap of the wavefunctions for scalar diquark is given by
\bea 
G_A^{u(S)}(Q^2)&=&\int\frac{d^2\bfp dx}{16\pi^3}~x\Big[\psi^{+(u)\dagger}_+(x,\bfp^\prime)\psi^{+(u)}_+(x,\bfp)\nonumber\\
&&-\psi^{+(u)\dagger}_-(x,\bfp^\prime)\psi^{+(u)}_-(x,\bfp)\Big],
\eea
and for vector diquark
\bea 
G_A^{\nu(A)}(Q^2)&=&\int\frac{d^2\bfp dx}{16\pi^3}~x\Big[\psi^{+(\nu)\dagger}_{++}(x,\bfp^\prime)\psi^{+(\nu)}_{++}(x,\bfp)\nonumber\\
&&-\psi^{+(\nu)\dagger}_{-+}(x,\bfp^\prime)\psi^{+(\nu)}_{-+}(x,\bfp)\nonumber\\
&&+\psi^{+(\nu)\dagger}_{+0}(x,\bfp^\prime)\psi^{+(\nu)}_{+0}(x,\bfp)\nonumber\\
&&-\psi^{+(\nu)\dagger}_{-0}(x,\bfp^\prime)\psi^{+(\nu)}_{-0}(x,\bfp)\Big],
\eea
where $\bfp^\prime = \bfp+(1-x)\bfq$.  In the SU(4) structure, depending on different flavors (struck quark) the axial form factor is written in terms of scalar and vector diquarks as
\bea 
G^u_{A}(Q^2)&=&C^2_S G_A^{u(S)}(Q^2) + C^2_V G_A^{u(V)}(Q^2),\label{Gfu}\\
G^d_{A}(Q^2)&=&C^2_{VV} G_A^{d(VV)}(Q^2).\label{Gfd}
\eea
Using the LFWFs given in Eqs.(\ref{LFWF_S},\ref{LFWF_Vp},\ref{LFWF_Vm}), 
the explicit calculation gives
\bea 
G_A^{u(S)}(Q^2)&=&N^2_S  G_1^{(u)}(Q^2),\\G_A^{u(V)}(Q^2)&=& (\frac{1}{3} N^{(u)2}_0 - \frac{2}{3} N^{(u)2}_1) G_1^{(u)}(Q^2), \\
G_A^{d(VV)}(Q^2)&=&(\frac{1}{3} N^{(d)2}_0 - \frac{2}{3} N^{(d)2}_1) G_1^{(d)}(Q^2), 
\eea
where $G_1^{(\nu)}(Q^2)$ are given by 
\bea 
&& G_1^{(\nu)}(Q^2)=\int dx \Big[x^{2a^{\nu}_1+1}(1-x)^{2b^{\nu}_1+1}\frac{1}{\delta^\nu}\nonumber\\
&&-x^{2a^{\nu}_2-1}(1-x)^{2b^{\nu}_2+3} \frac{\kappa^2}{(\delta^\nu)^2 M^2\log(1/x)} \nonumber\\
&&\times \Big(1-\delta^\nu\frac{Q^2}{4\kappa^2} \log(1/x)\Big)\Big]\exp\Big[-\delta^\nu\frac{Q^2}{4\kappa^2} \log(1/x)\Big].\label{G1}
\eea
\section{Soft-wall AdS/QCD model}
The explicit expressions for up and down quark GPDs in soft-wall AdS/QCD model can be represented as\cite{Sharma2014}
\begin{eqnarray}
H^q(x,Q^2)&=& \sum_\tau c_\tau q(x,\tau) x^{Q^2/4 \kappa^2},\nonumber\\
E^q(x,Q^2)&=&\sum _\tau c_\tau e^q(x,\tau) x^{Q^2/4 \kappa^2},
\end{eqnarray}
where, the quark distribution functions $q(x,\tau)$ and $e^q(x,\tau)$ can be written as
\begin{eqnarray}
q(x,\tau)&=& \alpha_1^q \gamma_1(x,\tau)+\alpha_2^q(x,\tau) \gamma_2(x,\tau)+\alpha_3^q \gamma_3(x,\tau),\nonumber\\
e^q(x,\tau)&=& \alpha_3^q \gamma_4(x,\tau).
\end{eqnarray}
The flavor coupling parameters $\alpha_i^q$ and $\gamma_i(x)$ are obtained from Ref. \cite{Sharma2014} and
\begin{eqnarray}
\gamma_1(x,\tau)&=&\frac{-1}{2}(1-2\tau+x \tau)(1-x)^{\tau-2},\nonumber\\
\gamma_2(x,\tau)&=&\frac{1}{2}(1-x \tau)(1-x)^{\tau-2},\nonumber\\
\gamma_3(x,\tau)&=&(1-3 \ x \ \tau+ x^2 \tau+x^2 \tau^2)(1-x)^{\tau-2},\nonumber\\
\gamma_4(x,\tau)&=& \frac{2 M_n}{\kappa}\tau \sqrt{\tau-1}(1-x)^{\tau-1}.
\end{eqnarray}
Here we have considered the three leading order dimensions $(\tau=3,4,5)$ which include the quarks, antiquarks and gluons.  

\begin{thebibliography}{99}
\bibitem{Diehl2003} M. Diehl, Phys. Rep. {\bf 388}, 41 (2003).
\bibitem{Goeke2001} K. Goeke, M. Polyakov and M. Vanderhaeghen, Prog. Part. Nucl. Phys. {\bf 47}, 401 (2001).
\bibitem{Adloffetal.2000a} C. Adloff {\it et. al.}, Eur.\  Phys.\  J.\  C {\bf 13}, 371 (2000).
\bibitem{adloff} C. Adloff {\it et. al.}, Phys. Lett. B {\bf 517}, 47 (2001).
\bibitem{Breitwegetal.1999} J. Breitweg {\it et. al.}, Eur.\  Phys.\  J.\  C {\bf 6}, 603 (1999).
\bibitem{Chekanov2003} S. Chekanov {\it et. al.}, Phys. Lett. B {\bf 573} 46 (2003).
\bibitem{Airapetian2001} A. Airapetian {\it et. al.}, Phys.\  Rev.\  Lett. {\bf 87}, 182001 (2001).
\bibitem{Stepanyan2001} S. Stepanyan {\it et. al.}, Phys.\  Rev.\  Lett. {\bf 87}, 182002 (2001).
\bibitem{dHoseN.2004} N. dHose, E. Burtin, P. A. M. Guichon and J. Marroncle, Eur. Phys. Jou. A {\bf 19}, 47 (2004).
\bibitem{AccardiA.2016} A. Accardi {\it et. al.}, Eur. Phys. Jou. A {\bf 52}, 268 (2016).
\bibitem{Pasquini2005b} B. Pasquini, M. Pincetti and S. Boffi, Phys.\  Rev.\  D {\bf 72}, 094029 (2005).
\bibitem{Pasquini2006} B. Pasquini and S. Boffi, Phys.\  Rev.\  D {\bf 73}, 094001 (2006).
\bibitem{Dahiya2008a} H. Dahiya and A. Mukherjee, Phys.\  Rev.\  D  {\bf 77}, 045032 (2008).
\bibitem{Frederico2009} T. Frederico, E. Pace, B. Pasquini and G. Salmè, Phys.\  Rev.\  D {\bf 80}, 054021 (2009).
\bibitem{Mukherjee2013} A. Mukherjee, S. Nair amd V. K. Ojha, Phys. Lett. B {\bf 721}, 284 (2013).
\bibitem{Maji:2015vsa} T. Maji, C. Mondal, D. Chakrabarti and O. V. Teryaev, JHEP {\bf 01}, 165 (2016).
\bibitem{Goldstein2011a} G. R. Goldstein, J. O. G. Hernadez and S. Liuti, Phys.\  Rev.\  D {\bf 84}, 034007 (2011).
\bibitem{Goldstein2015} G. R. Goldstein, J. O. G. Hernadez and S. Liuti, Phys.\  Rev.\  D {\bf 91}, 114013 (2015).
\bibitem{neetika2016} N. Sharma, Eur. Phys. Jou. A {\bf 52}, 338 (2016).
\bibitem{Sachs1962} R. G. Sachs, Phys. Rev. {\bf 126}, 2256 (1962).
\bibitem{Radyushkin1997} A. V. Radyushkin, Phys.\  Rev.\  D {\bf 56}, 5524 (1997).
\bibitem{Brodsky2001} S. J. Brodsky, D. S. Hwang, B.-Q. Ma and I. Schmidt, Nucl. Phys. B {\bf 593}, 311 (2001).
\bibitem{Kobzarev1962} I. Y. Kobzarev and L. B. Okun, Jou. Exp. Theor. Phys. {\bf 16}, 1343 (1962).
\bibitem{Kobsarev1970} I. Kobsarev and V. Zakharov, Annals of Physics {\bf 60}, 448 (1970).
\bibitem{Ji1997a} X. Ji, Phys.\  Rev.\  Lett. {\bf 78}, 610 (1997).
\bibitem{Ji1998} X. Ji, Phys.\  Rev.\  D {\bf 58}, 056003 (1998).
\bibitem{kumar2014a} N. Kumar and H. Dahiya, Mod. Phys. Lett. A {\bf 29}, 1450118 (2014).
\bibitem{Brodsky2008} S. J. Brodsky and G. F. de Tèramond, Phys.\  Rev.\  D {\bf 78}, 025032 (2008).
\bibitem{Abidin2009} Z. Abidin and C. E. Carlson, Phys.\  Rev.\  D {\bf 79} 115003 (2009).
\bibitem{Abidin2008a} Z. Abidin and C. E. Carlson, Phys.\  Rev.\  D {\bf 77} 115021 (2008).
\bibitem{Abidin2008b} Z. Abidin and C. E. Carlson, Phys.\  Rev.\  D {\bf 77} 095007 (2008).
\bibitem{Mondal2016a} C. Mondal, Eur.\  Phys.\  J.\  C {\bf 76}, 74 (2016).
\bibitem{Chakrabarti2015} D. Chakrabarti, C. Mondal and A. Mukherjee, Phys.\  Rev.\  D {\bf 91} 114026 (2015).
\bibitem{Selyugin2009} O. V. Selyugin and O. V. Teryaev, Phys.\  Rev.\  D {\bf 79}, 033003 (2009).
\bibitem{Ji2012} X. Ji., X. Xiong and F. Yuan, Phys. Lett. B {\bf 717}, 214 (2012).
\bibitem{Ji2012a}  X. Ji., X. Xiong and F. Yuan, Phys.\  Rev.\  Lett.  {\bf 109}, 152005 (2012).
\bibitem{Ji2013}  X. Ji., X. Xiong and F. Yuan, Phys.\  Rev.\  Lett.  {\bf 111}, 039103 (2012).
\bibitem{Leader2013} E. Leader and C. Lorcè, Phys.\  Rev.\  Lett. {\bf 111} 039101 (2013).
\bibitem{Harindranath2013} A. Harindranath, R. Kundu, A . Mukherjee and R. Ratabole, Phys.\  Rev.\  Lett. {\bf 111}, 039102 (2013).
\bibitem{Harindranath2014} A. Harindranath, R. Kundu, A . Mukherjee and R. Ratabole, Phys. Lett. B {\bf 728}, 63 (2014).
\bibitem{Abidin2008} Z. Abidin and C. E. Carlson, Phys.\  Rev.\  D {\bf 78} 071502 (2008).
\bibitem{Mondal2016b} C. Mondal, N.Kumar, H. Dahiya and D. Chakrabarti, Phys.\  Rev.\  D {\bf 94}, 074028 (2016).
\bibitem{Polyakov2003} M. Polyakov, Phys. Lett. B {\bf 555}, 57 (2003).
\bibitem{Goeke2007} K. Goeke {\it et. al.}, Phys.\  Rev.\  D {\bf 75}, 094021 (2007).
\bibitem{Adhikari2016} L. Adhikari and M. Burkardt, Phys.\  Rev.\  D {\bf 94}, 114021 (2016).
\bibitem{Leader2014} E. Leader and C.Lorcè, Phys. Rep. {\bf 541}, 163 (2014).
\bibitem{LiuKeh-Fei2016} K.-F. Liu and C.Lorcè, Eur. Phys. A {\bf 52}, 160 (2016).
\bibitem{Lorce2017} C. Lorcè, L. Mantovani and B. Pasquini, Phys. Lett. B {\bf 776}, 38 (2018).
\bibitem{Jakob1997} R. Jakob, P.Mulders and J. Rodrigues, Nucl. Phys. A {\bf 626}, 937 (1997).
\bibitem{Bacchetta2008} A. Bacchetta, F. Conti and M. Radici, Phys.\  Rev.\  D {\bf 78}, 074010 (2008).
\bibitem{Maji2016} T. Maji and D. Chakrabarti, Phys.\  Rev.\  D {\bf 94}, 094020 (2016).
\bibitem{Brodsky2008a} S. J. Brodsky and G. F. Tèramond, Phys.\  Rev.\  D. {\bf 77}, 056007 (2008).
\bibitem{Chakrabarti2017} D. Chakrabarti, T. Maji, C. Mondal and A. Mukherjee, Phys.\  Rev.\  D {\bf 95}, 074028 (2017).
\bibitem{Maji2017a} T. Maji and D. Chakrabarti, Phys.\  Rev.\  D {\bf 95}, 074009 (2017).
\bibitem{Mondal2017} C. Mondal, Eur.\  Phys.\  J.\  C {\bf 77}, 640 (2017).
\bibitem{Jaffe:2004ph} R. L. Jaffe, Phys. Rep. {\bf 409}, 1 (2005).
\bibitem{Jaffe:2005zz} Nucl. Phys. Proc. Suppl. {\bf 142}, 343 (2005).
\bibitem{Wilczek:2004im} F. Wilczek, {\it From fields to strings: Circumnavigating theoretical physics. Ian Koga memorial collection (3- volume set),} (World Scientific, Singapore, 2004), pp. 77-93.
\bibitem{Selem2006} A. Selem and F. Wilczek,  {\it Ringberg Workshop on New Trends in HERA physics 2005: Ringberg Castle, Tegernsee, Germany, October 2-7, 2005 }(World Scientific, Singapore, 2006), pp. 337-356.
\bibitem{Santopinto:2004hw} E. Santopinto, Phys. Rev. C {\bf 72}, 022201 (2005).
\bibitem{Forkel:2008un} H. Forkel and E. Klempt, Phys. Lett. B {\bf 679}, 77 (2009).
\bibitem{Cloet:2008reb} I. C. Cloet {\it et. al.}. Few Body Syst. {\bf 46}, 1 (2009).
\bibitem{Anisovich:2010wxy} A. V. Anisovich {\it et. al.}, Int. Jou. Mod. Phys. A {\bf 25}, 2965 (2010)[ Int. Jou. Mod. Phys. A {\bf 25}, 3155 (2010)]
\bibitem{Ferreti:2011zy} J. Ferreti, A. Vasallo and E. Santopinto, Phys. Rev. C {\bf 83}, 065204 (2011).
\bibitem{DeSanctis:2011zz} M. De Sanctis, J. Ferretti, E. Santopinto, and A. Vassallo,
Phys. Rev. C {\bf 84}, 055201 (2011).
\bibitem{Galata:2012xt} G. Galata and E. Santopinto, Phys. Rev. C {\bf 86}, 045202
(2012).
\bibitem{DeSanctis:2014ria} M. De Sanctis, J. Ferretti, E. Santopinto, and A. Vassallo,
Eur. Phys. Jou. A{\bf 52}, 121 (2016).
\bibitem{Eichmann:2011vu} G. Eichmann, Phys.\  Rev.\  D {\bf 84}, 014014 (2011).
\bibitem{MartinA.D.2009}. A. D. Martinand W. J. Stirling, R.S. Thorne, and G.Watt,
Eur. Phys. Jou. C {\bf 63}, 189 (2009).
\bibitem{Lepage1980} G. P. Lepage and S. J. Brodsky, Phys.\  Rev.\  D {\bf 22}, 2157
(1980).
\bibitem{Ellis2009} J. Ellis, D. S. Hwang, and A. Kotzinian, Phys.\  Rev.\  D {\bf 80},
074033 (2009).
\bibitem{Gutsche2014} T. Gutsche, V. E. Lyubovitskij, I. Schmidt, and A. Vega,
Phys.\  Rev.\  D {\bf 89}, 054033 (2014).
\bibitem{Gutsche2015} T. Gutsche, V. E. Lyubovitskij, I. Schmidt, and A. Vega,
Phys.\  Rev.\  D {\bf 92}, 019902 (2015).
\bibitem{Chakrabarti2013} D. Chakrabarti and C. Mondal, Phys.\  Rev.\  D {\bf 88}, 073006
(2013).
\bibitem{Sharma2014} N. Sharma, Phys.\  Rev.\  D {\bf 90}, 095024 (2014).
\bibitem{Miller2007} G. A. Miller, Phys.\  Rev.\  Lett. {\bf 99}, 112001 (2007).
\bibitem{Carlson2008} C. E. Carlson and M. Vanderhaeghen, Phys.\  Rev.\  Lett.
{\bf 100}, 032004 (2008).
\bibitem{Mondal2016} C. Mondal, Phys.\  Rev.\  D {\bf 94}, 073001 (2016).
\bibitem{Chakrabarti2014}. D. Chakrabarti and C. Mondal, Eur.\  Phys.\  J.\  C {\bf 74}, 2962 (2014).
\bibitem{MondalChandan2015} C. Mondal and D. Chakrabarti, Eur.\  Phys.\  J.\ C {\bf 75}, 261 (2015).
\bibitem{Kumar2014} N. Kumar and H. Dahiya, Phys.\  Rev.\  D {\bf 90}, 094030 (2014).\end{thebibliography}

  
\end{document}